\documentclass[aps,prd,reprint,twocolumn,superscriptaddress,longbibliography,nofootinbib,floatfix,showpacs]{revtex4-1}
\usepackage{epsfig}
\usepackage{amsmath,amssymb,amsfonts}
\usepackage{hyperref}
\usepackage{mathrsfs}
\usepackage{bbm}
\usepackage{slashed}
\usepackage{graphicx}
\usepackage{verbatim}

\usepackage{bm}

\usepackage[usenames]{color}

\definecolor{darkgreen}{rgb}{0.2,0.6,0}

\newcommand{\be}{\begin{equation}}
\newcommand{\ee}{\end{equation}}
\newcommand{\bw}{\begin{widetext}}
\newcommand{\ew}{\end{widetext}}
\newcommand{\bi}{\begin{itemize}}
\newcommand{\ei}{\end{itemize}}
\newcommand{\ud}{\mathrm{d}}

\newcommand{\LCm}{{\scriptscriptstyle -}} \newcommand{\LCp}{{\scriptscriptstyle +}}
\newcommand{\LCpm}{{\scriptscriptstyle \pm}}
\newcommand{\LCmp}{{\scriptscriptstyle \mp}}

\newcommand{\LCperp}{{\scriptscriptstyle \perp}}

\newcommand{\LCpara}{{\scriptscriptstyle \parallel}}

\usepackage[T1]{fontenc} \usepackage[latin1]{inputenc}

\begin{document}

\title{Worldline instantons for nonlinear Breit-Wheeler pair production and Compton scattering}

\author{Gianluca Degli Esposti}
\email{g.degli-esposti@hzdr.de}
\affiliation{Helmholtz-Zentrum Dresden-Rossendorf, Bautzner Landstra{\ss}e 400, 01328 Dresden, Germany}

\author{Greger Torgrimsson}
\email{greger.torgrimsson@umu.se}
\affiliation{Department of Physics, Ume{\aa} University, SE-901 87 Ume{\aa}, Sweden}
\affiliation{Helmholtz-Zentrum Dresden-Rossendorf, Bautzner Landstra{\ss}e 400, 01328 Dresden, Germany}

\begin{abstract}

Worldline instantons have previously been used to study the probability of Schwinger pair production (both the exponential and pre-exponential parts) and photon-stimulated pair production (the exponential part). Previous studies obtained the pair-production probability on the probability level by using unitarity, i.e. the imaginary part of the effective action for Schwinger pair production or the imaginary part of the polarization tensor for photon-stimulated pair production. The corresponding instantons are closed loops in the complex plane.
Here we show how to use instantons on the amplitude level, which means open instanton lines with start and end points representing fermions at asymptotic times. The amplitude is amputated with LSZ using, in general, field-dependent asymptotic states.
We show how to use this formalism for photon-stimulated/Breit-Wheeler pair production and nonlinear Compton scattering.

\end{abstract}
\maketitle

\section{Introduction}

Particle production by a weak field can be studied with saddle-point methods giving a probability that to leading order scales as $P=(\text{prefactor})\exp(-\text{exponent}/E)$, where $E$ is the field strength. For example, for a constant electric field one has\footnote{We absorb $e$ into the field, $eE\to E$, and use units with $c=\hbar=m_e=1$.}~\cite{Sauter:1931zz,Schwinger:1951nm}
\be
\text{Schwinger}:\qquad  P=\dots\exp\left\{-\frac{\pi}{E}\right\}
\ee
or for a time-dependent electric field one finds in general (see e.g.~\cite{Dunne:2005sx,Dunne:2006st})
\be
\text{time-dependent}: \qquad P=\dots\exp\left\{-\frac{F(\gamma)}{E}\right\} \;,
\ee
where $F$ is some function which depends on the pulse shape, $\gamma=\omega/E$ and $\omega$ is some characteristic frequency. For $\gamma\to e^\LCm e^\LCp$ in a constant electric field one has~\cite{Dunne:2009gi}
\be
P=\dots\exp\left\{-\frac{2}{E}\left([1+p^2]\text{arctan}\left[\frac{1}{p}\right]-p\right)\right\} \;,
\ee
where $p=\chi_\gamma/(2E)$ and $\chi_\gamma=\sqrt{-(Fk)^2}$. For $\gamma\to e^\LCm e^\LCp$ in a plane-wave Sauter pulse we have
\be
P=\dots\exp\left\{-\frac{4a_0}{\chi}[(1+a_0^2)\text{arccot}(a_0)-a_0]\right\} \;
\ee
where\footnote{In papers on pair production in a time-dependent electric field it is more common to use $\gamma$, while papers on processes in plane waves usually use $a_0$.} $a_0=E/\omega=1/\gamma$. Similar results, in fact with the same $a_0$ dependence in the exponent, hold for e.g. nonlinear Compton scattering $e^\LCm\to e^\LCm\gamma$~\cite{Dinu:2018efz} or trident pair production $e^\LCm\to e^\LCm e^\LCm e^\LCp$~\cite{Dinu:2017uoj}. There are of course many other examples. These results have been obtained e.g. using WKB approximations or the saddle-point method for approximating integrals that represent the exact result. 

Another semiclassical method is to use worldline instantons~\cite{Affleck:1981bma,Dunne:2005sx,Dunne:2006st}. The worldline formalism uses proper-time~\cite{FockProperTime,Schwinger:1951nm} and path integrals~\cite{Feynman:1948ur,Feynman:1950ir,Feynman:1951gn} to represents amplitudes or probabilities in terms of path integrals over particle trajectories, and a worldline instanton is a saddle point for such a path integral and is determined as a solution of the Lorentz force equation. Since we are interested in ``tunneling'' processes\footnote{By this we simply mean processes with probabilities that have exponential scaling.}, the instantons are necessarily complex.
It was initially used in~\cite{Affleck:1981bma} in order to study the probability for Schwinger pair production for a constant field but to all orders\footnote{Note that the zeroth order in $\alpha$ still contains all orders in $E$ (recall that we have absorbed $eE\to E$).} in $\alpha$. It was later realized~\cite{Dunne:2005sx,Dunne:2006st} that the worldline instanton formalism can also be useful to study Schwinger pair production by inhomogeneous fields. Although analytical results can only be obtained for certain simple fields (e.g. 1D electric fields depending only on one coordinate), the instanton approach offers a powerful method for fields depending on more than one coordinate~\cite{Dunne:2006ur,Dumlu:2015paa,Torgrimsson:2017cyb,Schneider:2018huk}. In fact, a numerical code was presented in~\cite{Schneider:2018huk} which allows us to study general fields depending on all space-time coordinates. This motivates us to develop the worldline-instanton formalism to other processes in strong fields. 

Our focus is on the worldline {\it instanton} formalism, which gives a saddle-point approximation for more complicated (and in principle general) field shapes. We note, though, that the worldline formalism~\cite{Strassler:1992zr,Schubert:2001he} has also been used to obtain exact results for various photon amplitudes in constant fields~\cite{Shaisultanov:1995tm,Schubert:2000yt} and general plane-wave background fields~\cite{Ilderton:2016qpj,Edwards:2021vhg}, open fermion lines in constant fields~\cite{McKeon:1994hd,Ahmad:2016vvw,Ahmadiniaz:2017rrk}, and Schwinger pair production for a class of fields for which the locally-constant-field (LCF) approximation is exact~\cite{Ilderton:2014mla}. The worldline integrals for various processes have also been evaluated numerically with a Monte Carlo method~\cite{Gies:2001zp,Gies:2005bz,Gies:2011he}.

In what could now be called the standard worldline-instanton approach~\cite{Dunne:2005sx,Dunne:2006st}, the Schwinger-pair-production probability is obtained from the imaginary part the effective action. The probability of pair production by a (single) photon in an electric field has been obtained in~\cite{Monin:2010qj,Satunin:2013an,Torgrimsson:2016ant} from the imaginary part of the polarization tensor. In all these cases the pair production probability $P$ is obtained by appealing to unitarity, which gives $P$ as the imaginary part of a single dressed fermion loop, with either no photons (Schwinger mechanism) or two photons ($\gamma\to e^\LCp e^\LCm$) attached. The fermion loop is represented in the worldline formalism as a path integral over closed worldline loops. Thus, in the standard approach the worldline instantons are closed loops, which are periodic in all 4 coordinates, $x^\mu(\tau_{\rm start})=x^\mu(\tau_{\rm end})$, where $\tau$ is the proper time. Although the starting point is the effective action (vacuum to vacuum amplitude) or the polarization tensor (photon to photon amplitude), when taking their imaginary part one is effectively working on the probability level, because the imaginary part gives directly the probability without having to take the absolute value squared. 

In this paper we will show how to use worldline instantons on the amplitude level. The starting point is a worldline representation that directly gives the amplitude of the considered process, rather than indirectly via the optical theorem. We are interested in processes with fermions in the asymptotic states. Apart from pair production (either spontaneous/Schwinger or stimulated/Breit-Wheeler $\gamma\to e^\LCp e^\LCm$), we are also interested in e.g. nonlinear Compton scattering $e^\LCm\to e^\LCm\gamma$. Compton scattering might perhaps not usually be thought of as a process with exponential scaling, but if the emitted photon has high energy then it has the same type of exponential scaling as nonlinear Breit-Wheeler. Having fermions in the asymptotic states means that we have open instanton trajectories, $x^\mu(\tau_{\rm start})\ne x^\mu(\tau_{\rm end})$. For Compton scattering $x^\mu(\tau\sim\tau_{\rm start})$ describes the electron motion before it enters the background field, at a complex space-time point $x^\mu(\tau_\gamma)$ a photon is emitted, and $x^\mu(\tau\sim\tau_{\rm end})$ describes the electron after it has left the field (we will also consider e.g. constant fields which is always present). For pair production $x^\mu(\tau\sim\tau_{\rm start})$ describes a positron at late times/in the final state, and $x^\mu(\tau\sim\tau_{\rm end})$ is an electron in the final state. So, in this case the instanton line starts in the future as a positron, moves backwards in time, enters the region with the background field, tunnels, and then moves forward in time. This agrees of course with the Stueckelberg-Feynman interpretation of positrons~\cite{StueckelbergWorldlineFigure,Feynman:1949hz}. Open worldlines have been used to study Schwinger pair production in constant electric fields in~\cite{Barut:1989mc,Rajeev:2021zae}.

At very high energies there are other semi-classical methods~\cite{BaierKatkovMethod,DiPiazza:2016maj} that can be used to study e.g. Breit-Wheeler pair production in general space-time dependent fields. With the worldline instanton methods we do not have to assume high energies, but instead we have to assume that we are in a regime where the probability has an exponential scaling. These different methods therefore complement each other.

This paper is organized as follows. In Sec.~\ref{Worldline instantons and LSZ} we briefly introduce the main ingredients. In~Sec.~\ref{Exponential part} we consider the exponential part of the probability of nonlinear Breit-Wheeler. In Sec.~\ref{Sauter pulse} we consider a Sauter pulse as an example and to illustrate explicitly the instanton solution. In Sec.~\ref{WKB solution from worldline instanton} we show how to calculate the pre-exponential factor, by deriving the WKB solution using Gelfand-Yaglom. In Sec.~\ref{Spectrum of pair production by a time-dependent field} we use the same method to calculate the pre-exponential factor of the momentum spectrum of pairs produced spontaneously in a general time-dependent, linearly-polarized electric field. In Sec.~\ref{preBWsection} we show how to use this method for nonlinear Breit-Wheeler in a non-constant field, which vanishes asymptotically, which is an example of application of Gelfand-Yaglom to a case where the instanton has a kink. In Sec.~\ref{Prefactor for nonlinear Compton} we apply the same method to obtain the pre-exponential factor of nonlinear Compton in a time-dependent electric field. Finally, in Appendix~\ref{prefactor constant field} we calculate the prefactor for nonlinear Breit-Wheeler in a constant electric field, which is an example where the asymptotic fermion states are non-trivial rather than just plane waves.

\section{Worldline instantons and LSZ}\label{Worldline instantons and LSZ}

The amplitude $M$ is obtained by amputating the fermion propagator using the LSZ approach (see e.g.~\cite{ItzyksonZuber}), either with a manifestly Lorentz invariant form
\begin{widetext}
\be\label{LSZ4pair}
M=\int\!\ud^4x\ud^4x'\; e^{ip_jx^j}\bar{u}_r^{(\rm asymp)}(t,{\bf p})(-i\slashed{\mathcal{D}}_x^\infty+m)S(x,x')(\overset{\leftarrow}{i\slashed{\mathcal{D}}_{x'}^\infty}+m)e^{ip'_jx'^j}v_{r'}^{(\rm asymp)}(t',{\bf p}') \;,
\ee
where $\mathcal{D}_\mu=\partial_\mu+iA_\mu$ and $\mathcal{D}_\mu^\infty=\partial_\mu+iA_\mu(t=\infty)$, or with
\be\label{LSZ3pair}
M=\lim_{t\to+\infty}\lim_{t'\to+\infty}\int\ud^3x\ud^3 x'\,e^{ip_jx^j}\bar{u}_r^{(\rm asymp)}(t,{\bf p})\gamma^0S(x,x')\gamma^0 e^{ip'_jx'^j}v_{r'}^{(\rm asymp)}(t',{\bf p}') \;,
\ee
where $S(x,x')$ is the dressed fermion propagator, which in an arbitrary background field can be expressed with the following worldline representation~\cite{Feynman:1951gn} (see~\cite{Fradkin:1991ci,Gies:2005ke,Ahmadiniaz:2020wlm,Corradini:2020prz} for different representations)
\be\label{propagatorWorldline}
S(x,x')=(i\slashed{\mathcal{D}}_x+m)\frac{1}{2}\int_0^\infty\!\ud T\int\limits_{q(0)=x'}^{q(1)=x}\mathcal{D}q\exp\left\{-i\left[\frac{Tm^2}{2}+\int_0^1\!\ud\tau\;\frac{\dot{q}^2}{2T}+A(q)\dot{q}\right]\right\}\mathcal{P}\exp\left\{-i\frac{T}{4}\int_0^1\sigma^{\mu\nu}F_{\mu\nu}\right\} \;,
\ee
\end{widetext}
where $\mathcal{P}$ means path ordering, i.e. ``time-ordering'' with respect to proper time $\tau$, and $\sigma^{\mu\nu}=\frac{i}{2}[\gamma^\mu,\gamma^\nu]$. Note that in the standard worldline-instanton approach one would work with periodic worldlines, $q(0)=q(1)$, but for the propagator one has $q(0)\ne q(1)$. Proper time has been normalized to $0<\tau<1$, so $T$ is the ``actual'' total proper time.

The explicitly Lorentz invariant form~\eqref{LSZ4pair} might be more common in the absence of a background field. One approach using~\eqref{LSZ4pair} would be to take the momenta $p_\mu$ and $p'_\mu$ temporarily off shell, so that one can perform partial integration to remove the derivatives acting on $S(x,x')$. However, then we would have two additional integrals (over $t$ and $t'$) as well as two nontrivial limits ($p^2\to1$ and $p'^2\to1$), while for~\eqref{LSZ3pair} we only have the limits $t,t'\to\infty$. Thus, at least for the time-dependent fields we have focused on here, we find it more convenient to use~\eqref{LSZ3pair}. This form has also been used in~\cite{Barut:1989mc} to obtain Schwinger pair production by a constant electric field.

In this paper we focus on fields that depend on time, but not on space. In this case, the asymptotic states, $u^{(\rm asymp)}$ and $v^{(\rm asymp)}$, can be obtained from the $t\to\pm\infty$ limit of the adiabatic/WKB approximations~\cite{Hebenstreit:2011pm,Hebenstreit:2010vz,Torgrimsson:2017pzs}. The (full) WKB approximations are at any time (not just asymptotic) given by
\be\label{UandV}
\begin{split}
	{\sf U}_r(t,{\bf q})&=(\gamma^0\pi_0+\gamma^i\pi_i+1){\sf G}^+(t,{\bf q})R_r \\
	{\sf V}_r(t,-{\bf q})&=(-\gamma^0\pi_0+\gamma^i\pi_i+1){\sf G}^-(t,{\bf q})R_r \;,
\end{split}
\ee
where $\pi_\LCperp=q_\LCperp$, $\pi_3(t)=q_3-A(t)$, $\pi_0=\sqrt{m_\LCperp^2+\pi_3^2(t)}$, $r=1,2$ denote two spin states, $\gamma^0\gamma^3 R_s=R_s$ and
\be\label{Gpmdef}
{\sf G}^\pm(t,{\bf q})=[2\pi_0(\pi_0\pm\pi_3)]^{-\frac{1}{2}}\exp\bigg[\mp i\int_{t_r}^t\!\ud t'\,\pi_0(t')\bigg] \;,
\ee
where $t_r$ is some arbitrary real constant.
If the electric field goes to zero asymptotically, then $u^{(\rm asymp)}$ and $v^{(\rm asymp)}$ are simple plane waves, but, since in general $A(-\infty)\ne A(\infty)$, the momentum depends on the asymptotic constant value of the gauge potential, $u^{(\rm asymp)}=\text{const.}e^{-i\pi_0(\infty)t}$. One can of course choose a gauge with $A(+\infty)=0$, but then we would in general have $A(-\infty)\ne0$, which would appear in e.g. Compton scattering where we have a fermion in the initial state. These plane-wave states for fields with $A'(\pm\infty)=0$ are of course easy to obtain without reference to the WKB solutions. However, for e.g. a constant electric field, the asymptotic states have a genuinely nontrivial dependence on the field, and in such a case it is convenient to obtain the asymptotic states from the $t\to\infty$ limit of ${\sf U}$ and ${\sf V}$. Note though, importantly, these nontrivial asymptotic states for e.g. a constant fields are still much simpler than the full, exact solutions to the Dirac equation, $u^{(\rm exact)}$ and $v^{(\rm exact)}$, which would involve parabolic cylinder functions for a constant field. This is an important point, because if it had been necessary to use $u^{(\rm exact)}$ and $v^{(\rm exact)}$, or even some approximation of these at finite times, then there would not really have been a point in using this worldline formalism. Fortunately, we only need the asymptotic states, so the only ``difficult'' field dependence is represented by the worldline path integral, which is given by~\eqref{propagatorWorldline} for any space-time dependent field. 

The worldline representation in~\eqref{propagatorWorldline} gives the fermion propagator in a completely arbitrary space-time dependent coherent background field. There is a common trick (see e.g.~\cite{Strassler:1992zr,McKeon:1994hd,Shaisultanov:1995tm,Schubert:2000yt,Schubert:2001he,Ahmad:2016vvw,Gies:2011he}) which allows us to include the absorption or emission of individual incoherent photons. For example, for nonlinear Breit-Wheeler, all we have to do is to replace $A_\mu\to A_\mu+\epsilon_\mu e^{-ikx}$ and select the term that is linear in $\epsilon_\mu$. For Compton scattering we would instead have $e^{+ikx}$, and the same trick also works for multiple photons.

\section{Exponential part}\label{Exponential part}

In order to introduce some of the main ideas, we will start with the exponential part of the probability. We return to calculate the prefactor in~\ref{WKB solution from worldline instanton}. We start with nonlinear Breit-Wheeler pair production, where an initial (incoherent) photon with momentum $k_\mu$ and polarization $\epsilon_\mu$ decays into an electron with momentum $p_\mu$ and a positron with momentum $p'_\mu$. We begin by making the replacement $A_\mu\to A_\mu+\epsilon_\mu e^{-ikx}$ in~\eqref{propagatorWorldline} and select the term linear in $\epsilon_\mu$. This gives
\be\label{expPartIntegrand}
\begin{split}
M=&\lim_{t\to\infty}\lim_{t'\to\infty}\int\ud^3x\ud^3 x'\int_0^\infty\ud T\int_0^1\ud\sigma\int_{q(0)=x'}^{q(1)=x}\mathcal{D}q \\
&\dots\exp\left\{-i\left[\frac{Tm^2}{2}+\int_0^1\!\ud\tau\;\frac{\dot{q}^2}{2T}+A(q)\dot{q}+Jq\right]\right\} \;,
\end{split}
\ee
where the proper time $\sigma$ integral comes from selecting the term that is linear in $\epsilon_\mu$, the ellipses stand for the prefactor part of the integrand, and the ``current'' is given by
\be\label{Jdefinition}
J_\mu=k_\mu\delta(\tau-\sigma) \;,
\ee
so $\sigma$ is the proper time when the photon decays. 
The prefactor part of the integrand also includes
\be\label{pathOrderedExp}
\mathcal{P}\exp\left\{-i\frac{T}{4}\int_0^1\sigma^{\mu\nu}F_{\mu\nu}\right\} \;,
\ee
because, even though it is given by an exponential, it is slowly varying, i.e. after a suitable rescaling of the integration variables the exponential written out in~\eqref{expPartIntegrand} scales as $\exp[i(\text{scalar part})/E]$, where $E\ll1$ is the field strength, while the exponent in~\eqref{pathOrderedExp} does not lead to terms with $1/E$ in the exponent. Thus, the saddle point, i.e. the worldline instanton, is the same in scalar and spinor QED. 

These paths can be thought of as electron lines, where the initial part has been bent into the future. So, the ``initial'' condition for $q(\tau)$ is a positron in the future, then the path goes backwards in time into the field, where it has a kink due to the photon absorption, turns and goes forward in time and ends as an electron in the future. 

Since we are considering a field which only depends on time, half of the spatial integrals give delta functions. We therefore change variables from $x'^j$ and $x^j$ to $\varphi^j=(x+x')^j/2$ and $\theta^j=(x-x')^j$, and then make a shift $q^j(\tau)\to\varphi^j+q^j(\tau)$. The ${\bm\varphi}$ integral gives $\delta^3({\bf p}+{\bf p}'-{\bf k})$. The boundary conditions for the spatial components of the path integral is now 
\be\label{thetaBoundary}
{\bf q}(0)=-\frac{\bm\theta}{2} 
\qquad
{\bf q}(1)=\frac{\bm\theta}{2} \;.
\ee

We will perform all the nontrivial integrals with the saddle-point method. In principle, one can perform them in any order, and in a future paper, where we plan to develop a numerical code using discretized worldlines, one would perform them all together and obtain the prefactor by calculating the determinant of a large Hessian matrix. However, for the time-dependent fields we consider here we can use the Gelfand-Yaglom method for calculating the path integral analytically, and for this reason it is better to perform the path integral first, while the exponent is still local in proper time. We therefore make a shift and a redefinition
\be\label{shiftRedifinition}
q_\mu(\tau)\to q_\mu(\tau)+\delta q(\tau) \;,
\ee
where from now on $q_\mu(\tau)$ is not an integration variable but a solution to the following Lorentz-force-like equation
\be\label{LorentzForceGeneral}
\ddot{q}^\mu=T(F^{\mu\nu}\dot{q}_\nu+J^\mu) \;,
\ee
which for the present case reduces to
\be\label{timeLorentz}
\ddot{q}_0=T(A'_j\dot{q}^j+J_0)
\ee 
and
\be\label{spaceLorentz}
\ddot{q}_j=T(-A'_j\dot{q}_0+J_i) \;,
\ee 
with boundary conditions~\eqref{thetaBoundary} for ${\bf q}$, $q_0(0)=t'$ and $q_0(1)=t$. 
The delta function in $J^\mu$ gives the instanton a kink, i.e. the instanton velocity is discontinuous at $\tau=\sigma$.
Since the boundary conditions for the original integration variable have been absorbed into the instanton, the new integration variable $\delta q_\mu(\tau)$ has Dirichlet boundary conditions, $\delta q_\mu(0)=\delta q_\mu(1)=0$. $q_\mu$ has been chosen to be a solution of this Lorentz-force equation such that the exponent contains no terms that are linear in $\delta q$, i.e. the instanton $q_\mu$ is a saddle point of the worldline path integral. 
The spatial part of the Lorentz-force equation~\eqref{spaceLorentz} gives immediately
\be\label{dotqA}
\dot{q}_i(\tau)=T(c_i+k_i\theta_{\tau\sigma}-A_i(q_0(\tau))) \;,
\ee
where $c_i$, $i=1,2,3$, are 3 constants, and $\theta_{\tau\sigma}=\theta(\tau-\sigma)$ is the step function.

An arbitrary variation of the instanton, $q_\mu\to q_\mu+\delta_a q_\mu$, leads to a variation of the instanton action
\be
\delta_a\int_0^1\!\ud\tau\;\frac{\dot{q}^2}{2T}+A\dot{q}+Jq=\left(\frac{\dot{q}}{T}+A\right)\delta_a q\bigg|_0^1 \;,
\ee
which is nonzero only if there is a variation of the end points. Since $q_0(0)=t'$ and $q_0(1)=t$ are not integration variables, only the spatial parts are relevant here. So, if we make a variation ${\bm\theta}\to{\bm\theta}+\delta{\bm\theta}$ then $\delta_{\bm\theta}{\bf q}(1)=\delta{\bm\theta}/2=-\delta_{\bm\theta}{\bf q}(0)$. Setting the linear variation in $\delta{\bm\theta}$ to zero gives 
\be\label{spatial01}
\left(\frac{\dot{q}_j}{T}+A_j\right)(0)+\left(\frac{\dot{q}_j}{T}+A_j\right)(1)=(p-p')_j \;.
\ee
This together with~\eqref{dotqA} implies $c_i=-p'_i$.

Setting the variation with respect to $\sigma$ to zero gives
\be\label{sigmaSaddle}
l\dot{q}(\sigma)=0 \;,
\ee
so the photon decays at a proper time when its 4-momentum is orthogonal to the instanton velocity.
This together with the Lorentz-force equation gives $\dot{q}\ddot{q}=\dot{q}(F\dot{q}+Jq)=0$, so $\dot{q}^2$ is independent of $\tau$. Note that without the photon kink, e.g. for instantons describing Schwinger pair production, the Lorentz-force equation would directly imply that $\dot{q}^2$ is a constant of motion, but here $\text{const.}=\dot{q}^2(\tau<\sigma)\ne\dot{q}^2(\tau>\sigma)=\text{const.}$ are in general two different constants before and after the photon absorption when the integration variables $\sigma$ is not equal to its saddle-point value.  

The saddle-point equation for $T$ is 
\be\label{TsaddleEq}
T^2=\int_0^1\dot{q}^2 \;,
\ee
which at the saddle point for $\sigma$ simplifies to
\be\label{onShellInstanton}
\dot{q}^2=T^2 \;.
\ee
This is an on-shell condition for the instanton. 
Substituting~\eqref{dotqA} into~\eqref{onShellInstanton} gives
\be\label{dotq0sol}
\begin{split}
\dot{q}_0(\tau)=&-\theta_{\sigma\tau}T\sqrt{1+({\bf p}'+{\bf A}(q_0(\tau)))^2}\\
&+\theta_{\tau\sigma}T\sqrt{1+({\bf p}-{\bf A}(q_0(\tau)))^2} \;.
\end{split}
\ee
Note that while $q$ is continuous, $\dot{q}$ cannot be continuous because there are delta functions in the Lorentz-force equation. 
From~\eqref{dotq0sol} it is also clear that the instanton initially moves backwards in time ($T$ turns out to have a large positive real part). 
But~\eqref{dotq0sol} does not automatically solve~\eqref{timeLorentz} at $\tau=\sigma$. Differentiating~\eqref{dotq0sol} and matching the resulting $\delta_{\tau\sigma}$ term with the one in~\eqref{timeLorentz} gives an additional condition\footnote{We use $\overset{!}{=}$ e.g. for a condition that is demanded in order to determine some parameter.}
\be\label{eqfortildet}
\sqrt{1+({\bf p}'+{\bf A}(\tilde{t}))^2}+\sqrt{1+({\bf p}-{\bf A}(\tilde{t}))^2}\overset{!}{=}\Omega \;,
\ee
where we have defined $\tilde{t}=q_0(\sigma)$ and $\Omega=k_0$. This equation gives us the time $\tilde{t}$ when the photon decays, and it turns out to be complex.

Now we have all the (implicit) saddle points, and the leading part of the probability is obtained by inserting the saddle points into the exponential. We first rewrite
\be\label{rewriteAdotq}
\int_0^1 A\dot{q}=p'_jx'^j+p_jx^j+\int_0^1\frac{\dot{q}_i^2}{T}-J_jq^j \;,
\ee
where we first used partial integration to obtain $q^j A'_j\dot{q}_0$, which is replaced using~\eqref{spaceLorentz}, and then a second partial integration for the $q_j\ddot{q}_j$ term. The terms in~\eqref{rewriteAdotq} all cancel against the other terms in the exponent. After this we are left with only one nontrivial $\tau$ integral, which we rewrite by changing variable from proper time $\tau$ to time $q_0(\tau)$,
\be
\int_0^1\frac{\dot{q}_0^2}{T}=\left(\int_0^\sigma+\int_\sigma^1\right)\dots=-\int_{t'}^{\tilde{t}}\ud t''\pi_{\LCm{\bf p}'}-\int_{t}^{\tilde{t}}\ud t''\pi_{\bf p} \;.
\ee
By comparing this with~\eqref{Gpmdef} we see that $t$ and $t'$ drop out from the exponent. 
We thus find
\be\label{finalExpGeneral}
\begin{split}
M\sim&\exp\bigg\{i\left[p'_jx'^j+p_jx^j+\int_{t_r}^{t'}\pi_{\LCm{\bf p}'}+\int_{t_r}^{t}\pi_{{\bf p}})\right] \\
&-i\left[\frac{Tm^2}{2}+\int_0^1\!\ud\tau\;\frac{\dot{q}^2}{2T}+A\dot{q}+Jq\right]\bigg\}\Bigg|_\text{saddle point} \\
=&\exp\left\{i\int_{t_r}^{\tilde{t}}\left[\pi_{\LCm{\bf p}'}+\pi_{{\bf p}}-\Omega\right]\right\} \;.
\end{split}
\ee 
This is the final result for the exponential part of the probability amplitude for a general time-dependent electric field. To evaluate it one just has to solve~\eqref{eqfortildet} to find the integration limit $\tilde{t}$ and then perform the time integral (the value of the lower integration limit $t_r\in\mathbb{R}$ is arbitrary). Eq.~\eqref{finalExpGeneral} agrees with the result in~\cite{Torgrimsson:2016ant}, which were obtained with either WKB or using unitarity (the optical theorem) to obtain the pair-production probability from the imaginary part of the photon polarization loop in the worldline representation. The main difference from the worldline derivation in~\cite{Torgrimsson:2016ant} is that~\cite{Torgrimsson:2016ant} considered a closed fermion loop, while here we have considered an open fermion loop. 

Note that we have obtained~\eqref{finalExpGeneral} without actually finding an explicit solution for the worldline instanton. All we needed in order to obtain this explicit final result are the implicit saddle-point/intanton equations. This is possible because the field only depends on one space-time coordinate. For a general space-time dependent field we will not be able to do this, we would have to actually find the instanton solution. However, considering a simple field can be very useful as a starting point for more general fields.

In~\cite{Schneider:2018huk} a numerical code was developed for obtaining the worldline instantons in a general space-time dependent electromagnetic field for the case of Schwinger pair production (i.e. pair production without the photon) (see also~\cite{Gould:2017fve}). The instanton is obtained by starting with the known, simple instanton in e.g. a constant field, and then the instanton in a general field is obtained by a numerical continuation, where the instanton is changed gradually by gradually changing the field from a constant to a general field.

The plan is to derive such a code also for photon-stimulated pair production, where the instanton in a general field is obtained from a numerical continuation of the simpler instanton in e.g. a purely time-dependent electric field. Thus, while we could obtain~\eqref{finalExpGeneral} without finding the instanton explicitly, it is nevertheless expected to be useful to go back and check what the instanton actually looks like. To do that we need to choose a field shape.

\section{Sauter pulse}\label{Sauter pulse} 
 
\subsection{Exponential part} 
 
Consider a linearly polarized electric field, $A_3=A(t)$. 
As an example, we consider a photon momentum ${\bf k}$ which is perpendicular to the electric field. The dominant contribution comes from a pair that shares the momentum equally between the electron and positron, i.e. ${\bf p}={\bf p}'={\bf k}/2$. Eq.~\eqref{eqfortildet} simplifies to
\be
\sqrt{m_\LCperp^2+A^2(\tilde{t})}=\frac{\Omega}{2} \qquad \to\qquad A(\tilde{t})=i\;,
\ee 
where
\be
m_\LCperp=\sqrt{1+\left(\frac{\Omega}{2}\right)^2}
\ee
is an ``effective'' mass, which comes from the fact that the absorbed photon not only provides energy (which enhances the probability) but also gives the pair momentum. At this point we need to choose a field shape. We consider a Sauter pulse
\be
A(t)=\frac{E}{\omega}\tanh(\omega t) \;.
\ee
We have
\be
\omega\tilde{t}=i\arctan\gamma \;,
\ee
where $\gamma=\omega/E$ is the Keldysh parameter (or $1/a_0$). Performing the integrals in~\eqref{finalExpGeneral} we find
\begin{widetext}
\be\label{expSauter}
|M|_\text{Sauter}^2\sim\exp\left\{-\frac{4}{E\gamma^2}\left(\sqrt{1+m_\LCperp^2\gamma^2}\text{arctan}\left[\frac{1}{p}\sqrt{1+m_\LCperp^2\gamma^2}\right]-\text{arctan}\frac{1}{p}-\gamma p\text{ arctan }\gamma\right)\right\} \;,
\ee
\end{widetext}
where $p=|{\bf p}|=\Omega/2$. This result interpolates between several different limits which can be compared with the literature.
Consider first the soft-photon limit,
\be
\lim_{\Omega\to0}\eqref{expSauter}=\exp\left\{-\frac{\pi}{E}\frac{2}{1+\sqrt{1+\gamma^2}}\right\} \;,
\ee
which agrees with the results in~\cite{PopovV35P659,Dunne:2006st,Dunne:2005sx} for the probability of pair production by a Sauter field without the additional photon. Consider next the slowly-varying-field limit,
\be
\lim_{\gamma\to0}\eqref{expSauter}=\exp\left\{-\frac{2}{E}\left([1+p^2]\text{arctan}\left[\frac{1}{p}\right]-p\right)\right\} \;,
\ee
which agrees with Eq.~(5) in~\cite{Dunne:2009gi} for pair production by a photon in a constant electric field. In the high-frequency limit we find
\be\label{perturbativeLimit}
\lim_{\gamma\gg1}\eqref{expSauter}=\exp\left\{-\frac{2\pi}{\omega}\left(\sqrt{1+p^2}-p\right)\right\} \;.
\ee 
Note that the electric-field strength $E$ has dropped out of the exponent in this limit. In fact, even though the saddle-point approximation of the pre-exponential factor breaks down in this limit, the exponent~\eqref{perturbativeLimit} is what one can expect from perturbative pair production: The Fourier transform of a Sauter pulse $\tilde{f}(w)$ scales at large Fourier frequencies $w$ as
\be\label{SauterFourier}
\tilde{f}(w)\sim\exp\left\{-\frac{\pi}{2}\frac{w}{\omega}\right\} \;.
\ee
It is possible to produce a pair by absorbing the high-energy photon plus one Fourier photon from the Sauter pulse if $w+\Omega\geq p_0+p'_0=2m_\LCperp$. The exponential suppression of the probability comes from the fact that the Fourier transform is exponentially suppressed at such high Fourier frequencies. Inserting the threshold value $w=2m_\LCperp-\Omega$ into~\eqref{SauterFourier} gives~\eqref{perturbativeLimit}. Compare with~\cite{Torgrimsson:2017pzs}.

The high-energy limit might be the most interesting limit. We find
\be\label{planeWaveLimit}
\lim_{\Omega\gg1}\eqref{expSauter}=\exp\left\{-\frac{4}{\gamma E\Omega}\left(\left[1+\frac{1}{\gamma^2}\right]\text{arccot}\left[\frac{1}{\gamma}\right]-\frac{1}{\gamma}\right)\right\} \;.
\ee
If we introduce $\chi:=\sqrt{-(F_{\mu\nu}k^\nu)^2}=E\Omega$ and write $a_0:=1/\gamma$ then we see that~\eqref{planeWaveLimit} is exactly the same as in Eq.~(60) in~\cite{Dinu:2017uoj} for trident ($e^\LCm\to e^\LCm e^\LCm e^\LCp$) pair production in a plane wave electromagnetic field, up to an overall process-dependent factor of $2nk/nP$ where $P_\mu$ is the momentum of an initial electron and $n_\mu$ is proportional to the wave vector of the laser ($n^2=0$).
The simplest thing to compare with would of course be the probability of Breit-Wheeler in a plane wave. However, we are not aware of such a result in the literature, so we have calculated it by applying the saddle-point method to the results in~\cite{Dinu:2019pau} (we will come back to this). The result agrees exactly with~\eqref{planeWaveLimit}.
The reason for this agreement is that a field effectively behaves as a plane wave in this limit because the field invariants are much smaller than $\chi$, so one can to leading order set ${\bf E}^2-{\bf B}^2=0$ and ${\bf E}\cdot{\bf B}=0$ which agrees with a plane wave field. Or one can make a Lorentz transformation to a frame where the photon energy is on the order of the electron mass, in order to make the frequency $\mathcal{O}(1)$ (this means $\Omega'\sim1$ in the new frame since we use units with $m_e=1$). In such a frame a general field looks like a plane wave. If in addition $a_0$ is large then the plane wave can be treated as a (locally) constant crossed field~\cite{Ritus1985}. In this double limit we have
\be
\lim_{\gamma\ll1}\lim_{\Omega\gg1}\eqref{expSauter}=\lim_{\gamma\ll1}\eqref{planeWaveLimit}=\exp\left\{-\frac{8}{3\chi}\right\} \;,
\ee
which is the well-known scaling for nonlinear Breit-Wheeler pair production in the constant-crossed-field approximation~\cite{Reiss62,Nikishov:1964zza}.
However, we see from~\eqref{planeWaveLimit} that the high-energy limit agrees with the plane-wave result in a larger regime, i.e. not just for large $a_0$ but also for $a_0\gtrsim1$. It is useful to see that our result~\eqref{expSauter} interpolates to this high-energy/plane-wave limit as this means that the instanton approach can be used also for fields that are closer to plane waves (e.g. single laser beam) rather than combination of two or more laser beams with significantly nonzero field invariants.

\subsection{Instanton}

To obtain the instanton we first calculate $T$. Note that the saddle-point equation~\eqref{TsaddleEq} which we obtained by varying $T$ only gives an implicit equation because the instanton $q$ depends on $T$. Instead we can obtain $T$ from
\be
\begin{split}
T=&T\left(\int_0^\sigma+\int_\sigma^1\right)\ud\tau=\left(\int_{\tilde{t}}^t+\int_{\tilde{t}}^{t'}\right)\frac{\ud\bar{t}}{\sqrt{m_\LCperp^2+A^2(\bar{t})}} \\
 =&\frac{1}{\sqrt{1+m_\LCperp^2\gamma^2}E}\bigg\{\text{arcsinh}\left[\frac{1}{m_\LCperp}\sqrt{m_\LCperp^2+\frac{1}{\gamma^2}}\sinh(\omega t)\right] \\
&+\text{arcsinh}\left[\frac{1}{m_\LCperp}\sqrt{m_\LCperp^2+\frac{1}{\gamma^2}}\sinh(\omega t')\right]\\
&-2i\,\text{arcsin}\left[\frac{1}{m_\LCperp}\sqrt{\frac{1+m_\LCperp^2\gamma^2}{1+\gamma^2}}\right]\bigg\}\;.
\end{split}
\ee
Similarly,
\be
\begin{split}
&T\sigma=T\int_0^\sigma\ud\tau\\
&=\frac{1}{\sqrt{1+m_\LCperp^2\gamma^2}E}\bigg\{\text{arcsinh}\left[\frac{1}{m_\LCperp}\sqrt{m_\LCperp^2+\frac{1}{\gamma^2}}\sinh(\omega t)\right]\\
&-i\,\text{arcsin}\left[\frac{1}{m_\LCperp}\sqrt{\frac{1+m_\LCperp^2\gamma^2}{1+\gamma^2}}\right]\bigg\}\;.
\end{split}
\ee
For asymptotic times $t,t'\gg1$ we have
\be
T\sim\frac{t+t'}{\pi_0(\infty)} \qquad \sigma\sim\frac{t}{t+t'} \qquad \pi_0(\infty)=\sqrt{m_\LCperp^2+\frac{1}{\gamma^2}} \;,
\ee
which is the proper time $T$ required for a positron to start at $t$ with asymptotic momentum $\pi(\infty)$, go back in time to a time period where the field is nonzero $q_0\sim1$, turn and go back to the future again, where it is an electron. Of course, even if $t$ and $t'$ are real and so the real part of $T$ (and $\sigma$) is much larger than the imaginary part, we cannot neglect the imaginary part because it is needed to tunnel. For $t=t'$ we have $\sigma=1/2$ exactly.

\begin{figure*}
\includegraphics[width=.32\linewidth]{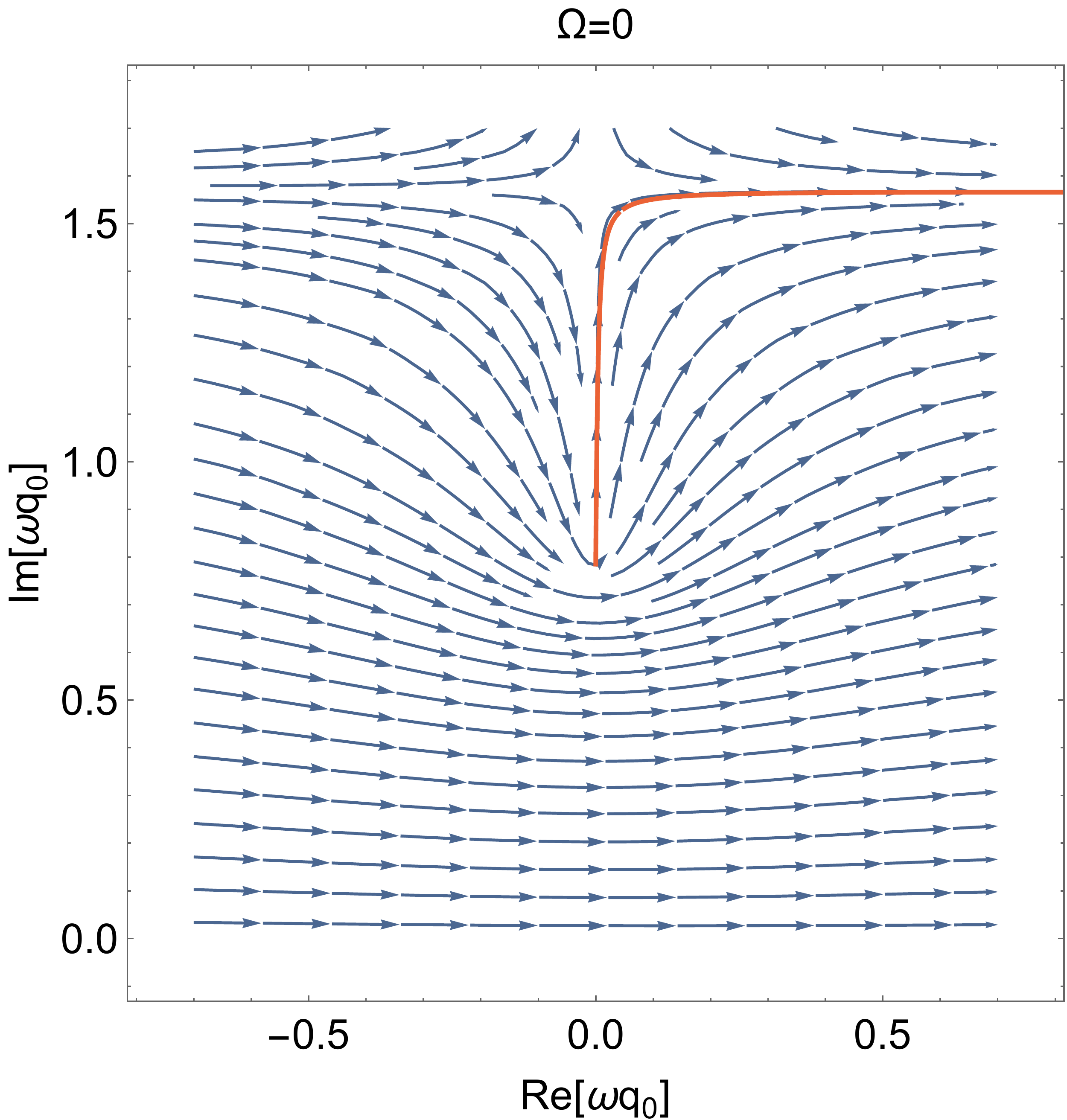}
\includegraphics[width=.32\linewidth]{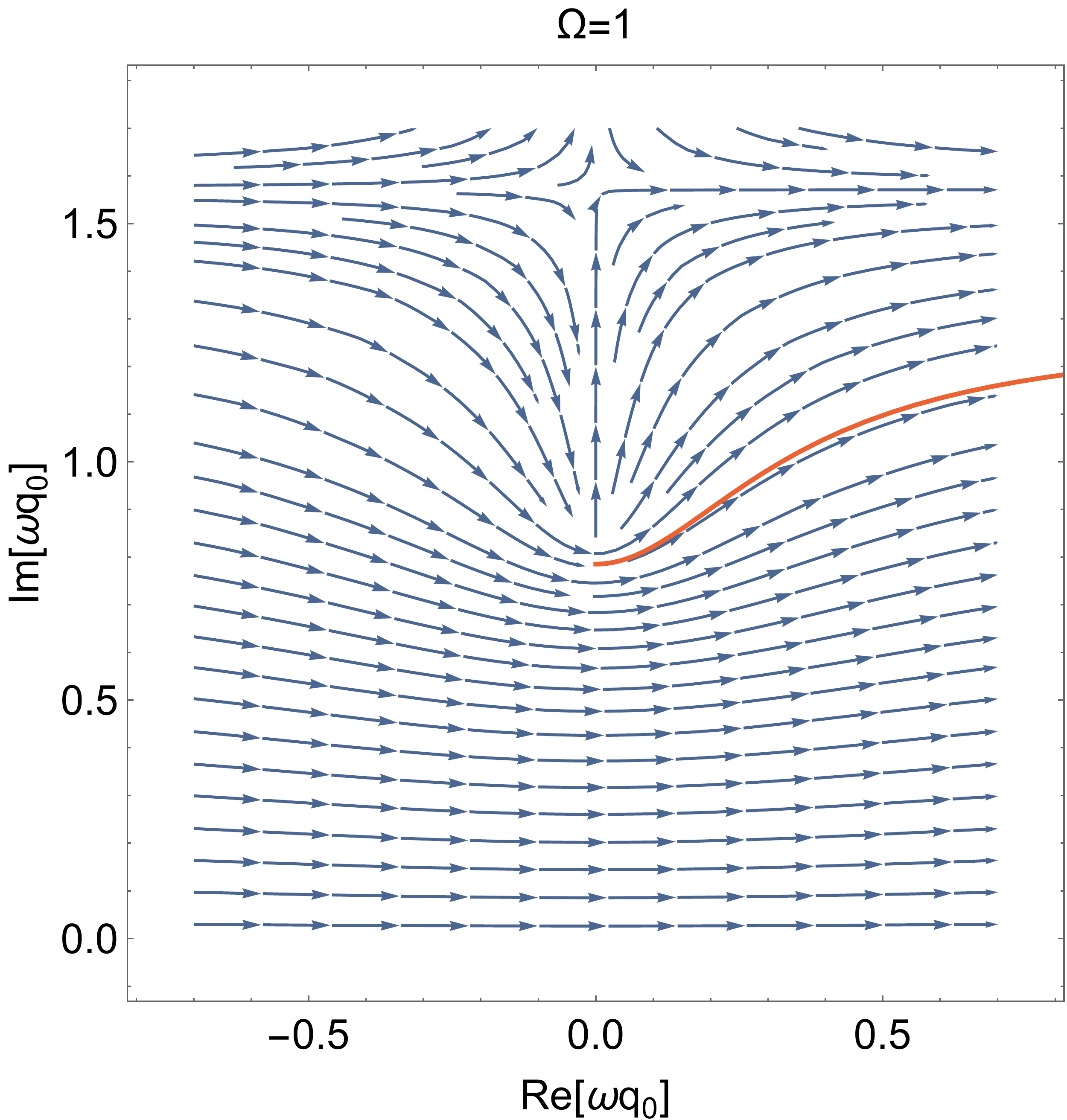}
\includegraphics[width=.32\linewidth]{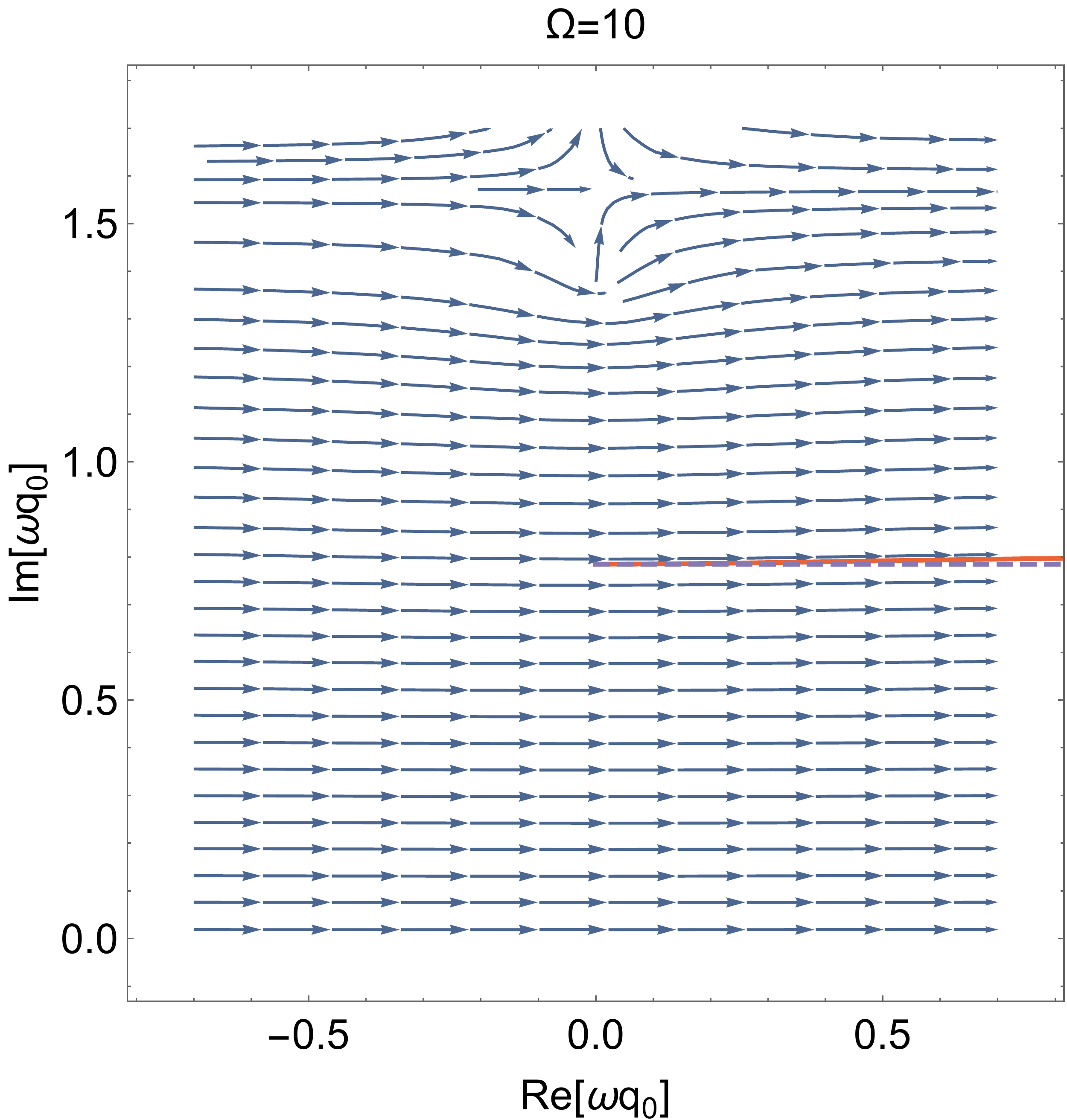}
\caption{Stream lines show the velocity (or energy rather) $\dot{q}_0=\pm T\sqrt{m_\LCperp^2+A^2(q_0)}$, cf.~\eqref{dotq0sol} with $p_\LCperp^2={p'_\LCperp}^2=(\Omega/2)^2$ and $p_3=p'_3=0$. The red solid lines show the analytical instanton solution~\eqref{SauterInstantonU}. The dashed line shows the high-energy approximation of the instanton~\eqref{highEnergyInstanton}. $\gamma=1$ in all plots. For $\Omega=1,10$, the instanton is plotted with proper time $u$ along the real axis, and $T$ has been chosen real for the stream lines. However, for $\Omega=0$ we have chosen $u=e^{i\vartheta}r$ with real $r$ and a small phase, $\vartheta$, in order to prevent the instanton from going into the pole of the Sauter pulse at $\omega q_0=i\pi/2$. For the stream plot we have similarly chosen $T=e^{i\vartheta}|T|$. The precise value of $\vartheta$ is not important, but for this particular plot we have chosen $\vartheta=-.001\pi$. Note that the stream arrows sometimes point in the opposite direction compared to the velocity of an actual trajectory. For example, for $u<0$ the plotted analytical instanton starts at $q_0\to+\infty$ and moves backwards to the turning point, i.e. opposite to direction of the plotted stream arrows, while for $u>0$ it turns back along the same line.}
\label{streamPlotsSauter}
\end{figure*}

Before presenting the explicit instanton solution, we will first derive a general expression for the final exponential expressed in terms of the instanton solution, so that we can check that the explicit instanton gives the correct result~\eqref{expSauter}. For a completely general space-time dependent field, the instanton is a solution to the Lorentz-force equation
\be
\frac{\ddot{q}_\mu}{T}=F_{\mu\nu}\dot{q}^\nu+J_\mu \;,
\ee
where $F_{\mu\nu}=\partial_\mu A_\nu-\partial_\nu A_\mu$. With two partial integrations we find
\be
\int_0^1A\dot{q}=\left(\frac{\dot{q}}{T}+A\right)q\bigg|_0^1+\int_0^1\left(-q^\mu\partial_\mu A_\nu\dot{q}^\nu-\frac{\dot{q}^2}{T}-Jq\right)
\ee
so
\be
\begin{split}
&-i\left[\frac{Tm^2}{2}+\int_0^1\!\ud\tau\;\frac{\dot{q}^2}{2T}+A\dot{q}+Jq\right]\\
&=-i\left(\frac{\dot{q}}{T}+A\right)q\bigg|_0^1+i\int_0^1q^\mu\partial_\mu A_\nu\dot{q}^\nu \;.
\end{split}
\ee
The boundary terms should cancel against the asymptotic states.
We change variables from $\tau$ to $u=T(\tau-\sigma)$ and take the limit $t,t'\to\infty$, which means $T\to\infty$, and find (we assume here that $A'(\pm\infty)=0$)
\be\label{finalExpGeneralField}
|M|^2\sim\exp\left\{2\text{Re }i\int_{-\infty}^\infty\!\ud u\; q^\mu\partial_\mu A_\nu\frac{\ud q^\nu}{\ud u}\right\} \;.
\ee
Note that $u$ is actually what one would usually call proper time, because instead of~\eqref{onShellInstanton} the on-shell condition for the instanton reads $(\ud q/\ud u)^2=1$. In fact, \eqref{finalExpGeneralField} is reparameterization invariant. 
We are of course free to make a contour deformation for proper time $u$; choosing a real or a complex contour changes of course the instanton path, but not the integral. 
Note also that~\eqref{finalExpGeneralField} only depends on the photon implicitly via the instanton solution, i.e.~\eqref{finalExpGeneralField} works for both photon-stimulated and spontaneous pair production.

We return now to the Sauter pulse. In terms of $u$ we have for $u>0$:
\be\label{SauterInstantonU}
\begin{split}
q_0(u)&=\frac{1}{\omega}\text{arcsinh}\bigg\{\frac{m_\LCperp\gamma}{\sqrt{1+m_\LCperp^2\gamma^2}}\\
&\times\sinh\left(U+i\,\text{arcsin}\left[\frac{1}{m_\LCperp}\sqrt{\frac{1+m_\LCperp^2\gamma^2}{1+\gamma^2}}\right]\right)\bigg\} \\
q_3(u)&=-\frac{1}{\sqrt{1+m_\LCperp^2\gamma^2}\omega}\bigg[\text{arcsinh}\bigg\{m_\LCperp\gamma\\
&\times\cosh\left(U+i\,\text{arcsin}\left[\frac{1}{m_\LCperp}\sqrt{\frac{1+m_\LCperp^2\gamma^2}{1+\gamma^2}}\right]\right)\bigg\}\\
&-\text{arcsinh}\left\{\gamma\sqrt{\frac{m_\LCperp^2-1}{1+\gamma^2}}\right\}\bigg] \;,
\end{split}
\ee
where 
\be\label{Ufromu}
U=\sqrt{1+m_\LCperp^2\gamma^2}Eu \;,
\ee
and the solution at $u<0$ is simply obtained from $q_0^{(u<0)}(u)=q_0^{(u>0)}(-u)$ and $q_3^{(u<0)}(u)=-q_3^{(u>0)}(-u)$. The solution in~\eqref{SauterInstantonU} is still exact, i.e. it is obtained by just changing variables from $\tau$ to $u$ without having to take any $T,t,t'\to\infty$ limit. By plugging~\eqref{SauterInstantonU} into~\eqref{finalExpGeneralField} and performing the integral numerically we can check that the instanton solution indeed gives the correct result~\eqref{expSauter}.

For a photon with high energy, i.e. in the limit where the time-dependent background field behaves as a plane wave, the instanton simplifies considerably, and the time component is just a straight line before and after the photon absorption,
\be\label{highEnergyInstanton}
\lim_{\Omega\gg1}q_0(U>0)=\frac{1}{\omega}(U+i\,\text{arctan}\,\gamma) \;.
\ee
The corresponding approximation for $\dot{q}_3$ can be obtained from~\eqref{dotqA} and is, up to a constant, just $A_3(q_0)$. It is straightforward to check that by inserting this leading high-energy approximation of the instanton into~\eqref{finalExpGeneralField} we obtain~\eqref{planeWaveLimit}. From~\eqref{Ufromu} we see that $\dot{q}_0=\ud q_0/\ud u\to\pm\Omega/2$, which is expected since the electron and positron share the energy of the absorbed photon equally and the $u<0$ and $u>0$ halves of the instanton correspond, respectively, to the positron and electron. However, even in this $\Omega\gg1$ limit the appropriate integration variable in~\eqref{finalExpGeneralField} is $U$, i.e. the $\Omega\gg1$ limit is obtained by expanding the integrand with $U$ rather than $u$ as independent of $\Omega$, so knowing that $\dot{q}_0\to\pm\Omega/2$ is not enough.     
 
The instanton and its high-energy approximation are shown in Fig.~\ref{streamPlotsSauter}. Note that nowhere in this study has it been necessary to rotate to euclidean time. Fig.~\ref{streamPlotsSauter} also suggests that such a rotation would not be helpful here, as the time component of the instanton $q_0$ is neither purely imaginary nor real.

\subsection{Instantons in a plane-wave background}

We have seen that the probability of Breit-Wheeler in a time dependent electric field $E(t)$ reduces to the one in a plane-wave background when the photon frequency is very high. In this section we will show how to obtain the plane-wave result by working directly with the instanton in a plane wave. In the $E(t)$ case we have chosen the field to point along the $z$ axis and focused on photons with perpendicular momentum, $k_\mu=\Omega(1,1,0,0)$, so to show that such a high-energy photon effectively ``sees'' a plane wave we would boost along the $x$ axis. Hence, for comparison, we will now choose a plane wave traveling along the $x$ axis and with polarization along the $z$ axis, with nonzero component $A_3=a_3(\phi)=a_0f(\phi)$, where $a_0=E/\omega$ and $\phi=Kx=\omega(t+x)$. We assume for simplicity a symmetric field $a(-\phi)=-a(\phi)$. Lightfront components are given by $v^\LCpm=2v_\LCmp=v^0\pm v^1$ and $v_\LCperp=\{v_2,v_3\}$ for an arbitrary vector $v_\mu$. We will show that the exponent in~\eqref{finalExpGeneralField} gives the correct result. We therefore use proper time $u$ rather than $\tau$, 
\be\label{uLorentzForce}
\frac{\ud^2q_\mu}{\ud u^2}=F_{\mu\nu}\frac{\ud q^\nu}{\ud u}+k_\mu\delta(u) \;,
\ee
where $F_{\mu\nu}=K_\mu a'_\nu-a'_\mu K_{\nu}$. As is well known, the Lorentz-force equation without the current term ($k_\mu\delta(u)$) has a simple exact solution for a general plane wave. The solution is still simple with the current term,
\be
\begin{split}
&u>0: \qquad \phi=\tilde{\phi}+Kp u \qquad \frac{\ud q_\LCperp}{\ud u}=(p-a)_\LCperp\\
&u<0: \qquad \phi=\tilde{\phi}-Kp u \qquad \frac{\ud q_\LCperp}{\ud u}=(-p-a)_\LCperp
\end{split}
\ee
where $\phi=Kq$. The remaining component can be obtained from the on-shell condition,
\be\label{qplusOnShell}
\frac{\ud q_\LCp}{\ud u}=\frac{1+(\ud q_\LCperp/\ud u)^2}{4\ud q_\LCm/\ud u} \;.
\ee 
Here we have assumed that the electron and positron share the absorbed photon momentum equally, $p_{\LCm,\LCperp}=p'_{\LCm,\LCperp}=k_{\LCm,\LCperp}/2$. The ``turning point'' $\tilde{\phi}$ is determined by~\eqref{sigmaSaddle} and ensures that the $\delta(u)$ term in $\ud^2_u q_\LCp$ agrees with~\eqref{uLorentzForce},
\be
a_\LCperp^2(\tilde{\phi})\overset{!}{=}-1 \;.
\ee
Since $a(-\phi)=-a(\phi)$ this implies that $\tilde{\phi}=iz$ is imaginary, $z>0$. Inserting this instanton into~\eqref{finalExpGeneralField}, changing variable from $u$ to $\phi$ and performing a partial integration gives
\be\label{planeWaveExpGenf}
\begin{split}
|M|^2\sim&\exp\left\{-\frac{4}{Kl}\text{Re }i\left(-\tilde{\phi}+\int_{\tilde{\phi}}^\infty\ud\phi\, a_\LCperp^2\right)\right\}\\
=&\exp\left\{-\frac{4z}{Kl}\left(1+a_0^2\frac{1}{2iz}\int_{-iz}^{iz}\ud\phi\,f^2(\phi)\right)\right\} \;,
\end{split}
\ee
where in the second step we have chosen a $\phi$ contour that first goes down from $\tilde{\phi}$ to $0$ and then from $0$ to $\infty$ along the real axis; the second part cancels because it is a pure phase. \eqref{planeWaveExpGenf} agrees with Eq.~(93) in~\cite{Dinu:2018efz}\footnote{That equation is for nonlinear Compton, but the only difference is that $r_{10}$ there should be replaced with $r_{\rm BW}$, which is $r_{\rm BW}\to4$ for the saddle-point value of the longitudinal momentum $Kp$.}, which was obtained using the Volkov solutions. \eqref{planeWaveExpGenf} works for a general symmetric field. For a Sauter pulse we recover~\eqref{planeWaveLimit}.

\section{WKB solution from worldline instanton}\label{WKB solution from worldline instanton}

So far we have focused on the exponential part of the probability. We now turn to the problem of calculating the pre-exponential part using the worldline formalism and the Gelfand-Yaglom method.

For one-dimensional fields one can often do the calculations conveniently using the WKB approximations of the Dirac equation. As a first application of open worldline instantons we therefore start by deriving these WKB solutions.

We begin with the path integral. Expanding around the instanton gives a zeroth order which we have already discussed, a linear term that vanishes, and a quadratic term on the form
\be\label{pathQuadLambda}
\exp\left\{-\frac{i}{2T}\int_0^1\begin{pmatrix}
	\delta t & \delta z 
\end{pmatrix}
\Lambda
\begin{pmatrix}
	\delta t \\ \delta z
\end{pmatrix}
\right\} \;,
\ee
where
\be
\Lambda=\begin{pmatrix}
	-\partial^2+TA''\dot{z} & TA'\partial \\
	-\partial TA' & \partial^2 
\end{pmatrix} \;,
\ee
where $\partial=\partial/\partial\tau$.
The path integral over $\delta t, \delta z$ gives us the funtional determinant of $\Lambda$. We will calculate it using the Gelfand-Yaglom method, which was applied to closed instantons for spontaneous pair production in~\cite{Dunne:2006st}. For this we need to find the two solutions, $\phi^{(1)}$ and $\phi^{(2)}$, of
\be\label{LambdaEq}
\Lambda\phi=0 
\qquad
\phi(0)=0
\ee
with
\be\label{initialdphi}
\dot{\phi}^{(1)}(0)=\begin{pmatrix} 1\\0\end{pmatrix}
\qquad
\dot{\phi}^{(2)}(0)=\begin{pmatrix} 0\\1\end{pmatrix} \;.
\ee
The determinant is given by
\be\label{detLambdaIstart}
\det\Lambda=(\phi^{(1)}_1\phi^{(2)}_2-\phi^{(1)}_2\phi^{(2)}_1)|_{\tau=1} \;.
\ee
The path integral over $\delta q_\mu$ gives the free part (this includes both $\delta t$, $\delta z$ and $\delta q_\LCperp$)
\be
\int_{\delta q(0)=0}^{\delta q(1)=0}\mathcal{D}\delta q\exp\left\{-i\int_0^1\frac{\delta\dot{q}^2}{2T}\right\} =\frac{1}{(2\pi T)^2}
\ee
times $\sqrt{\det\Lambda_{\rm free}}/\sqrt{\det\Lambda}$, but with our normalization we have $\phi^{(1)}_{\rm free}=\{\tau,0\}$ and $\phi^{(2)}_{\rm free}=\{0,\tau\}$, so $\det\Lambda_{\rm free}=1$.

To solve~\eqref{LambdaEq} we make the ansatz
\be\label{phiAnsatz}
\phi(\tau)=h(\tau)\begin{pmatrix}
	\dot{t} \\ \dot{z}
\end{pmatrix}
+d(\tau)\begin{pmatrix}
	0 \\ 1
\end{pmatrix} \;,
\ee
where $h$ and $d$ are now the functions to be determined.
One of the two components of $\Lambda\phi=0$ gives
\be
\partial[\dot{d}+\partial(h\dot{z})-TA'h\dot{t}]=0 \;,
\ee
which integrates to
\be
d=c_1\tau-h\dot{z}+\int_0^\tau TA'h\dot{t} \;,
\ee
where $c_1$ is a constant. For the second component we have
\be
\partial(\dot{h}\dot{t}^2-c_1TA)=0 \;,
\ee 
which leads to
\be
\phi_1=h\dot{t}=\dot{t}\int_0^\tau\frac{\ud\tau'}{\dot{t}^2}[c_1TA+c_2] \;,
\ee
where $c_2$ is an additional constant, and so
\be
\begin{split}
\phi_2&=h\dot{z}+d=c_1\tau+\int_0^\tau TA'\phi_1 \\
&=c_1\tau+\int_0^\tau\frac{\ud\tau'}{\dot{t}^2}[c_1TA+c_2]T[A_\tau-A] \;.
\end{split}
\ee
We determine $c_1$ and $c_2$ from the initial conditions~\eqref{initialdphi} 
\be\label{c1c2fromInitial}
c_1^{(1)}=0 \qquad c_2^{(1)}=\dot{t}_0
\qquad
c_1^{(2)}=1 \qquad c_2^{(2)}=-TA(t_0) \;.
\ee
We have three integrals
\be
\begin{split}
I_0&=\int_0^1\frac{1}{\dot{t}^2}
\qquad
I_1=\int_0^1\frac{TA}{\dot{t}^2}
\\
I_2&=\int_0^1\frac{(TA)^2}{\dot{t}^2}=1+2Tp_3I_1-T^2(m_\LCperp^2+p_3^2)I_0 \;.
\end{split}
\ee
where we have used
\be
\dot{t}=T\sqrt{m_\LCperp^2+(p_3-A(t))^2} \;.
\ee
We find
\be
\phi(1)=\begin{pmatrix}
	\dot{t}_1(c_1I_1+c_2I_0)
	 \\
	c_1[1+TA(t_1)I_1-I_2]+c_2[TA(t_1)I_0-I_1]
\end{pmatrix} \;.
\ee
The determinant~\eqref{detLambdaIstart} becomes
\be\label{detLambdaI}
\det\Lambda=\dot{t}_0\dot{t}_1(I_0+I_1^2-I_0I_2) \;.
\ee
For $t_1\to\infty$ we have
\be
I_0\to\frac{t_1}{T^3\pi_0^3(\infty)}
\qquad
I_1\to\frac{t_1A(\infty)}{T^2\pi_0^3(\infty)}
\qquad
I_2\to\frac{t_1A^2(\infty)}{T\pi_0^3(\infty)} \;.
\ee
Since $T$ is also large in this limit we can drop the terms with $I_1$ and $I_2$ and we find
\be
\det\Lambda\to\frac{t_1\pi_0(t_0)}{T\pi_0^2(\infty)}  \;.
\ee

Next we turn to the ordinary integrals. For $\det\Lambda$ we could use the final form of the instanton, but now we need the instanton as a function of general $T$ and start/end points. From the Lorentz-force equation we have
\be
\dot{q}^2=\text{constant}=:T^2a^2 \;,
\ee
which defines a constant $a$.
For the spatial components we have
\be
\dot{x}^\LCperp=Tc^\LCperp
\qquad
\dot{z}=T(c^3+A_3) \;,
\ee
where $c^j$ are three constants determined by the initial and final points,
\be
c^\LCperp=\frac{\Delta x^\LCperp}{T}
\qquad
c^3=\frac{\Delta z}{T}-\int_0^1\ud\tau\, A_3 \;,
\ee
where $\Delta x=x(1)-x(0)$.
We define
\be
G(a^2,c^j):=\int_{t_0}^{t_1}\ud t\sqrt{a^2+c_\LCperp^2+(c^3+A_3)^2} \;,
\ee
which gives a function for arbitrary arguments $a,c^j$. The actual values of $a,c^j$ in the instanton can now be determined from
\be\label{a2cjFromG}
G_0:=\frac{\partial}{\partial a^2}G\overset{!}{=}\frac{T}{2}
\qquad
G^j:=\frac{\partial}{\partial c_j}G\overset{!}{=}-\Delta x^j \;.
\ee
The instanton action can be expressed as $e^{-iS}$
\be\label{expa2cj}
S=\frac{T}{2}+\int_0^1\frac{\dot{x}^2}{2T}+A_3\dot{z}
=\frac{T}{2}(1-a^2)-c_j\Delta x_j+G \;.
\ee
We have
\be
\frac{\ud S}{\ud T}=\frac{1}{2}(1-a^2) \;,
\ee
where the terms with $\ud a^2/\ud T$ and $\ud c_j/\ud T$ cancel due to~\eqref{a2cjFromG}. Thus,
\be
a^2(T=T_{\rm saddle},\Delta x_j)=1 \;,
\ee
or
\be
T_{\rm saddle}(\Delta x_j)=2G_0(1,c^j) \;.
\ee
For the prefactor we also need the second derivative. To obtain this we first differentiate~\eqref{a2cjFromG} with respect to $T$, solve $\ud G_j/\ud T=0$ for $\ud c_j/\ud T$ in terms of $\ud a^2/\ud T$ and substitute into $\ud G_0/\ud T=1/2$, which gives
\be
\frac{\ud^2}{\ud T}S=-\frac{1}{4}\left(G_{00}-G_{0j}G^{-1}_{jk}G_{0k}\right)^{-1} \;,
\ee 
where $G_{00}=\left(\frac{\partial}{\partial a^2}\right)^2G$, $G_{0j}=\frac{\partial}{\partial a^2}\frac{\partial}{\partial c^j}G$ and $G^{-1}_{jk}$ is the $j,k$ element of the inverse of the matrix $G_{ij}=\frac{\partial}{\partial c^i}\frac{\partial}{\partial c^j}G$.
Thus, (up to an irrelevant phase) the contribution from the $T$ integral to the prefactor is give by
\be\label{TintG}
\int\ud T\to 2\sqrt{2\pi}\left(-G_{00}+G_{0j}G^{-1}_{jk}G_{0k}\right)^{1/2} \;,
\ee
by which we mean that the full result is obtained by making this replacement in addition to replacing $T\to T_{\rm saddle}$ in the integrand.
In the asymptotic limit we find
\be\label{TintGasymp}
\eqref{TintG}\to\sqrt{\frac{2\pi t_1}{\pi_0(\infty)}} \;.
\ee

Now we turn to the integral over $x^j=x^j(1)$. The exponential part of the integrand is given by
\be\label{xjexp}
i p_j x^j-iS=i(p_j x^j-c_j\Delta x^j-G) \;.
\ee
Upon differentiating with respect to $x^j$ we find that the terms with $\ud c_k/\ud x^j$ vanish due to~\eqref{a2cjFromG} and hence 
\be
\frac{\ud}{\ud x^j}\eqref{xjexp}=i(p_j-c_j) \;,
\ee
so
\be
c_j(x^k_{\rm saddle})=p_j
\ee
or from~\eqref{a2cjFromG}
\be
x^j_{\rm saddle}=x_j(0)-G_j(1,{\bf p}) \;.
\ee
By differentiating the second equation in~\eqref{a2cjFromG} we obtain
\be
\frac{\ud}{\ud x^j}\frac{\ud}{\ud x^k}\eqref{xjexp}=iG^{-1}_{jk} \;.
\ee
Thus, the spatial integrals give (up to an irrelevant phase)
\be\label{xjint}
\int\ud^3x\to(2\pi)^{3/2}\sqrt{\det G_{jk}} \;,
\ee
which in the asymptotic limit gives
\be\label{xjintAsymp}
\lim_{t_1\to\infty}\eqref{xjint}=(2\pi)^{3/2}\frac{t_1^{3/2}}{\pi_0^{5/2}(\infty)} \;.
\ee

The final exponent is now obtained from~\eqref{xjexp} and the term from the asymptotic state~\eqref{UandV},
\be
\eqref{xjexp}+i\int_{t_r}^{t_1}\pi_0=ip_j x_{(0)}^j+i\int_{t_r}^{t_0}\pi_0 \;,
\ee
which agrees with the exponential part of the WKB solution~\eqref{UandV}.

For the prefactor we find
\be
\frac{1}{(2\pi T)^2}\frac{1}{\sqrt{\det\Lambda}}\int\ud T\int\ud^3x\to\frac{1}{\sqrt{\pi_0(t_0)\pi_0(\infty)}} \;.
\ee
The spin factor simplifies
\be
\begin{split}
&\mathcal{P}\exp\left\{-\frac{iT}{4}\int_0^1\sigma^{\mu\nu}F_{\mu\nu}\right\} \\
&=\exp\left\{\frac{T}{2}\int_0^1\ud\tau A_3'(t)\gamma^0\gamma^3\right\} \\
&=\exp\left\{\frac{1}{2}\ln\left[\frac{\pi_0(\infty)-\pi_3(\infty)}{\pi_0-\pi_3}\right]\gamma^0\gamma^3\right\} \;,
\end{split}
\ee
where the integral was performed by changing variable from $\tau$ to $t$. The rest of the spinor part is given by $\bar{R}(\slashed{\pi}_\infty+1)\gamma^0(\slashed{\pi}_\infty+1)=2\pi^0_\infty\bar{R}(\slashed{\pi}_\infty+1)$. Putting everything together we finally find
\be
\frac{\bar{R}(\slashed{\pi}+1)}{\sqrt{2\pi_0(\pi_0+\pi_3)}}\exp\left\{ip_j x_{(0)}^j+i\int_{t_r}^{t_0}\!\ud t\,\pi_0\right\} \;,
\ee
which agrees exactly with the WKB solution $\bar{\sf U}$ in~\eqref{UandV}.

\section{Spectrum of spontaneous pair production by a time-dependent field}\label{Spectrum of pair production by a time-dependent field}

\begin{figure}
\includegraphics[width=\linewidth]{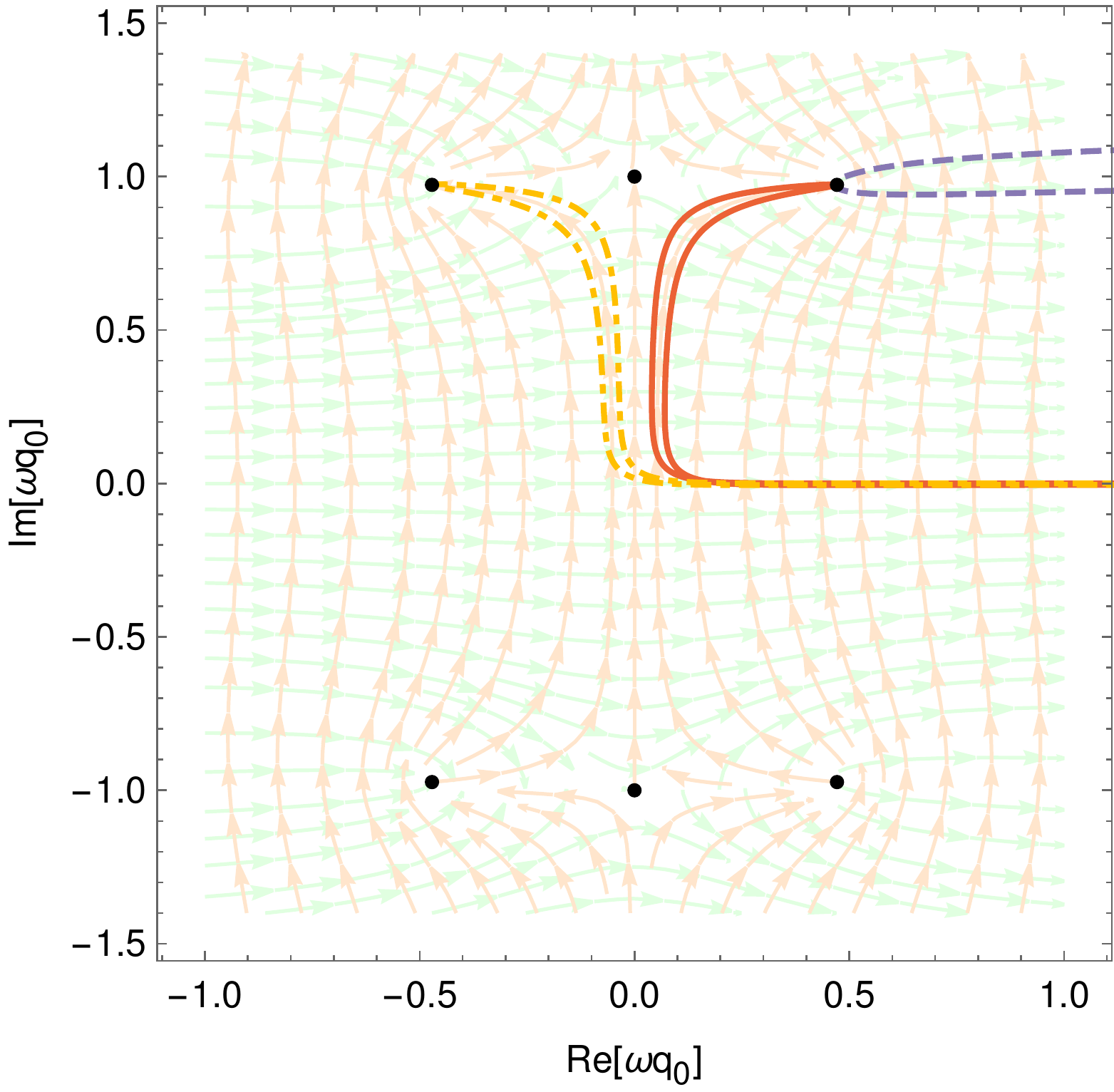}
\includegraphics[width=\linewidth]{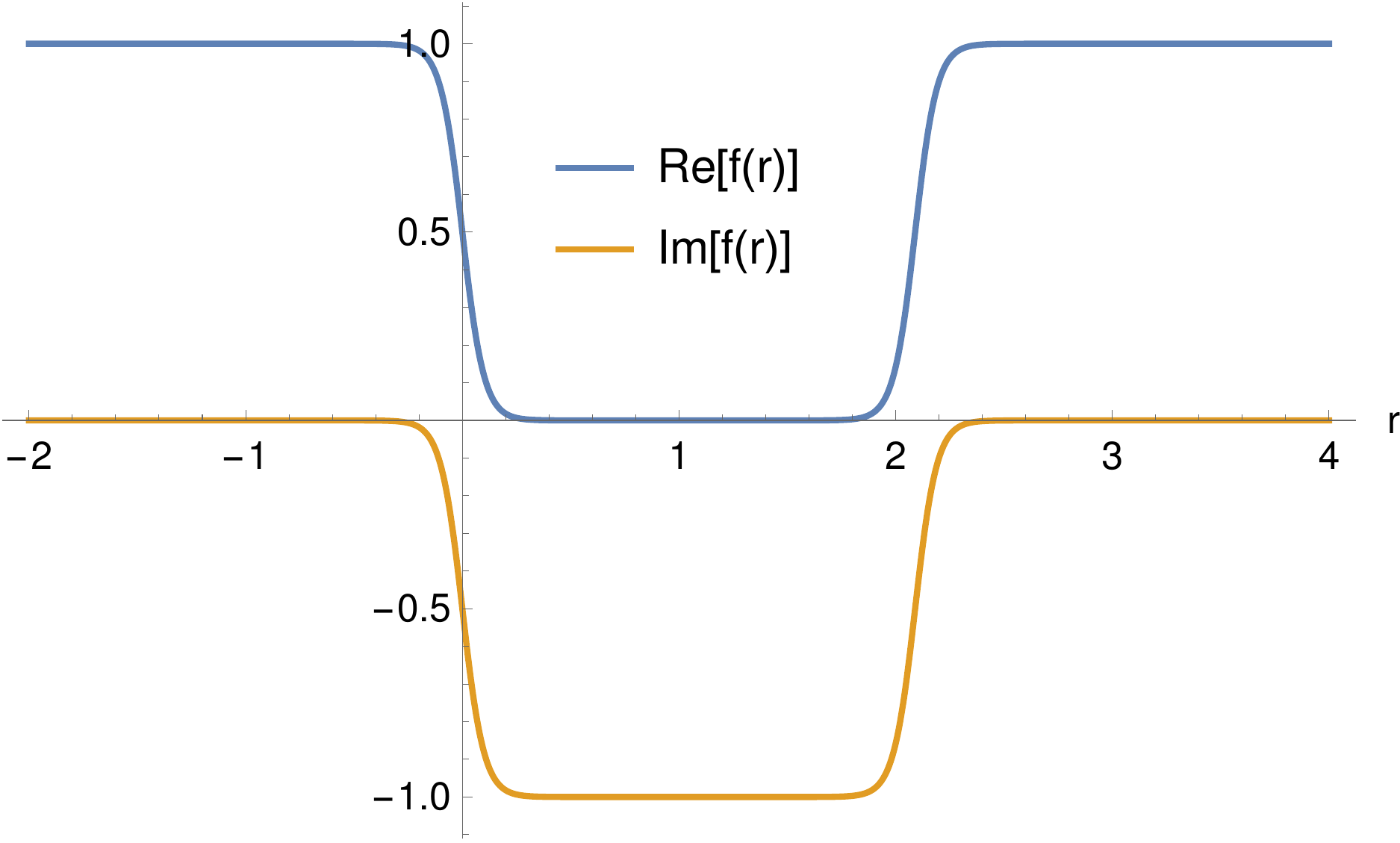}
\caption{The first plot shows instantons for the field in~\eqref{interferenceField}, with $\gamma=1$ and longitudinal momentum $p_3=.3$. The light green stream lines, i.e. those that on the real axis are parallel to the real axis, show~\eqref{streamSqrtReal}, while the light orange stream lines are orthogonal~\eqref{streamSqrtIm}. The black dots show the poles of the field at $\omega q_0=\pm i$ and the four turning points. The dashed purple line shows an example of an instanton that starts and ends at $\text{Re }q_0\to\infty$ and goes around one of the turning points, with proper time $u$ real throughout the trajectory. The solid red line shows an example of an instanton for which proper time follows a complex contour, specifically parameterized as in~\eqref{fWithTanh} with $W=.1$ and $L=2.09$. The yellow dot-dashed line shows a similar instanton that goes around the other (upper) turning point.}
\label{streamPlotInterferenceFig}
\end{figure}

\subsection{Worldline derivation}

A very similar calculation gives us the pair-production amplitude and so the spectrum. The $G$ function is now defined as
\be
G(a^2,c^j):=\int_0^1\ud\tau\frac{\dot{t}^2}{T}=2\int_{\tilde{t}}^{t_1}\ud t\sqrt{a^2+c_\LCperp^2+(c^3+A_3)^2} \;,
\ee
where we have set $t_0=t_1$ and $\tilde{t}=t(\sigma)$ is the (complex) time where $\dot{t}(\sigma)=0$, i.e. the turning point where the instanton stops moving backwards in time and starts moving forward.
Instead of~\eqref{detLambdaI} we find ($\dot{t}_1=-\dot{t}_0=\pi_0(\infty)$)
\be
\det\Lambda\to-\frac{t_0+t_1}{T\pi_0(\infty)} \;,
\ee
instead of~\eqref{TintGasymp} we have
\be
\eqref{TintG}\to\sqrt{\frac{2\pi (t_0+t_1)}{\pi_0(\infty)}} \;,
\ee
and similarly for~\eqref{xjintAsymp}. Thus (up to an irrelevant phase)
\be
\frac{1}{(2\pi T)^2}\frac{1}{\sqrt{\det\Lambda}}\int\ud T\int\ud^3x\to\frac{1}{\pi_0(\infty)} \;.
\ee
The other spatial integrals give the momentum conserving delta function, $\int\ud^3x'\to(2\pi)^3\delta^3({\bf p}+{\bf{p}'})$.
The spinor part is also similar to before, but with one important difference,
\be
\mathcal{P}\exp\left\{-\frac{iT}{4}\int_0^1\sigma^{\mu\nu}F_{\mu\nu}\right\}=\exp\left\{\ln\left(\frac{\varepsilon i m_\LCperp}{\pi_0+\pi_3}\right)\gamma^0\gamma^3\right\} \;,
\ee
where $\varepsilon=\pm1$ is determined by
\be
p_3-A_3(\tilde{t})=\varepsilon i m_\LCperp \;.
\ee
We find (up to an irrelevant phase)
\be\label{finalAmpSpectrum}
M=(2\pi)^3\delta^3({\bf p}+{\bf{p}'})R_{s'}^\dagger R_s \varepsilon \exp\left\{2i\int_{t_r}^{\tilde{t}}\pi_0\right\} \;,
\ee 
where $R_{s'}^\dagger R_s=\delta_{ss'}$ for $s,s'=1,2$.
For a field with only one maximum, like a Sauter pulse, we only have one saddle point and the sign $\varepsilon$ is irrelevant. However, for a field with multiple maxima/minima we have several saddle points and hence in the coherent sum of these we need to keep the relative sign given by $\varepsilon$. 

\subsection{Instantons}

The amplitude in~\eqref{finalAmpSpectrum} agrees with the result in~\cite{Dumlu:2011rr}, which was obtained by a saddle-point treatment of the Riccati equation. The momentum spectrum for time-dependent electric fields was also studied with worldline instantons in~\cite{Dumlu:2011cc}, but with an important difference: In~\cite{Dumlu:2011cc} a worldline representation was used for the effective action, which gives directly the probability rather than the amplitude. Those worldlines are all closed loops, i.e. $x^\mu(0)=x^\mu(1)$, while we are working with open worldlines. Working on the probability level increases the number of instantons one has to deal with. For example, for a field with one maximum and one minimum we have two instantons on the amplitude level but four on the probability level. 

As an example we consider
\be\label{interferenceField}
A_3(t)=\frac{1}{\gamma}\frac{1}{1+(\omega t)^2} \;.
\ee
In contrast to the Sauter pulse, here we have two turning points with $\text{Im}(t)>0$,
\be
\omega t_m=-\sqrt{\frac{1}{\gamma(p_3+i)}-1}
\ee
and,
\be
\omega t_p=\sqrt{\frac{1}{\gamma(p_3-i)}-1} \;.
\ee
For pair production we need instantons that start and end at $t\to+\infty$. By simply plotting the stream lines for
\be\label{streamSqrtReal}
\dot{q}_0=\pm|T|\sqrt{1+(p_3-A_3(q_0))^2}
\ee
we can immediately see instantons that wrap around the turning point $t_p$ with $\text{Re } t_p>0$ and $\text{Im } t_p>0$, see Fig.~\ref{streamPlotInterferenceFig}. Note that there is not just one unique instanton because the instanton does not actually have to go though the turning point, it just has to wrap around it and so one can continuously increase the distance between the instanton to the turning point (up to the closest pole or branch cut). In fact, the turning point is also a branch point, so one might expect that it can in some cases be numerically advantageous to choose instantons that do not go too close to the turning point.
We can also see instantons that wrap around the other turning point, $t_m$ with $\text{Re } t_m<0$ and $\text{Im } t_m>0$, but they start and end at $t\to-\infty$ and therefore do not satisfy the boundary conditions. The instantons that follow the stream lines of~\eqref{streamSqrtReal} have real proper time from start to end. 

To find instantons that go around $t_m$ but have the correct boundary conditions, i.e. starting and ending at $t\to+\infty$, we can use a complex contour for the proper time, i.e. use a complex einbein, cf.~\cite{Dumlu:2011cc}. We parameterize the complex proper time contour in terms of a real variable $r$. We denote the Jacobian for this change of variable  
\be
\frac{\ud u}{\ud r}=f(r) \;.
\ee
When $f$ is real the instanton follows the stream lines of~\eqref{streamSqrtReal}. When $f$ is imaginary the instanton instead follows
\be\label{streamSqrtIm}
\dot{q}_0=\pm i|T|\sqrt{1+(p_3-A_3(q_0))^2} \;,
\ee
i.e. stream lines that are orthogonal to~\eqref{streamSqrtReal}. As can be seen in Fig.~\ref{streamPlotInterferenceFig}, instantons that follow the stream lines of~\eqref{streamSqrtIm} can wrap around either $t_p$ or $t_m$, but those instantons also wrap around the turning points in the lower half complex plane and form closed loops. In other words, if we choose $f$ imaginary for the whole trajectory then we find no instantons that start and end at $t\to+\infty$. Instead, we can choose a $f$ that is sometimes real and sometimes imaginary. There is no unique choice of $f$. We have chosen
\be\label{fWithTanh}
\begin{split}
f(r)=&\frac{1}{2}\left(1-\tanh\left[\frac{r}{W}\right]\right)+\frac{1}{2}\left(1+\tanh\left[\frac{r-L}{W}\right]\right) \\
&-\frac{i}{4}\left(1+\tanh\left[\frac{r}{W}\right]\right)\left(1-\tanh\left[\frac{r-L}{W}\right]\right) \;,
\end{split}
\ee   
illustrated in Fig.~\ref{streamPlotInterferenceFig}. 
To find an instanton we can look at the stream lines in Fig.~\ref{streamPlotInterferenceFig} and select a point $t_0$ which we guess should be on the instanton line. Then we can guess a point $r_0$ which wound make the instanton go around either $t_p$ or $t_m$. By tuning these parameters as well as $L$ and $W$ can also guide $q_0$ so that it follows the real axis asymptotically (as $r\to\pm\infty$). Two such instantons are shown in Fig.~\ref{streamPlotInterferenceFig}. By inserting these numerical instanton solutions into (cf.~\eqref{finalExpGeneralField})   
\be\label{finalExpGeneralFieldInterference}
\exp\left\{i\int_{-\infty}^\infty\!\ud u\; q^\mu\partial_\mu A_\nu\frac{\ud q^\nu}{\ud u}\right\}
\ee
we find the same real part of the exponent as in~\eqref{finalAmpSpectrum} and the relative phase for the two contributions also agrees with~\eqref{finalAmpSpectrum}.

We have thus shown explicit examples of instantons in a case with interference. We emphasize again, though, that the instanton path for $q_0$ is really rather arbitrary since we have a great deal of freedom in choosing the complex proper time contour. Choosing $f$ with (smooth) step-function like behavior as in~\eqref{fWithTanh} is of course just one choice that seems convenient. This arbitrariness corresponds to the arbitrariness in choosing the complex contour for the $t$ integral in~\eqref{finalAmpSpectrum} or~\eqref{finalExpGeneral}; these $t$ integrals were obtained by changing integration variable from proper time $u$ to $t=q_0(u)$, but the resulting $t$ integral can of course be deformed in the complex plane without even having to think about instantons. Of course, this does not make the instanton path completely arbitrary. One can for example not deform the proper time contour to guide $q_0$ and $q_3$ separately. Also, the two instantons in Fig.~\ref{streamPlotInterferenceFig} which follow the real axis asymptotically are not equivalent; they give the same real part of the exponent in~\eqref{finalExpGeneralFieldInterference}, but different imaginary part.  
Moreover, here we have focused on purely time-dependent fields, but for fields that depend on more than one space-time coordinate one might not be able to rewrite the results as expression like~\eqref{finalAmpSpectrum} or~\eqref{finalExpGeneral} that no longer involves the instanton. In such cases it might not be as obvious what sort of arbitrariness in the instanton path one has. However, one always has the freedom to choose different proper time contours.

Using open worldlines is a more general approach, because we can reproduce results obtained with closed worldlines for the effective action, while e.g. Compton scattering cannot be treated with such closed lines.
It can also lead to further insights into the process since in the open worldline instantons we can see real, physical particles emerging at asymptotic times from complex paths during the tunneling at finite times.

\section{Prefactor for nonlinear Breit-Wheeler pair production in time-dependent electric fields}\label{preBWsection}

In this section we will calculate the prefactor for nonlinear Breit-Wheeler pair production in non-constant fields. 

\subsection{Worldline derivation}

Expanding around the instanton gives again~\eqref{pathQuadLambda}, and we again solve~\eqref{LambdaEq} using the ansatz in~\eqref{phiAnsatz}. One of the two components gives
\be
\partial[\dot{d}+\partial(h\dot{z})-TA'h\dot{t}]=0 \;,
\ee
which integrates to
\be
d=c_1\tau-h\dot{z}+\int_0^\tau TA'h\dot{t} \;,
\ee
where $c_1$ is a constant. For the second component we have to be careful with the kink at $\tau=\sigma$. So, we consider first $\tau\ne\sigma$, where the second component gives
\be
\partial(\dot{h}\dot{t}^2-c_1TA)=0 \;,
\ee 
which leads us to
\be\label{phi1BW}
\phi_1=h\dot{t}=\dot{t}\left\{\int_0^\tau\frac{\ud\tau'}{\dot{t}^2}[c_1TA+c_2+c_3\theta_{\tau'\sigma}]+c_4\theta_{\tau\sigma}\right\} \;,
\ee
where $c_2,c_3,c_4$ are three additional constants, and
\be\label{phi2BW}
\begin{split}
&\phi_2=h\dot{z}+d=c_1\tau+\int_0^\tau TA'\phi_1 \\
&=c_1\tau+c_4\theta_{\tau\sigma}T[A_\tau-A_\sigma] \\
&+\int_0^\tau\frac{\ud\tau'}{\dot{t}^2}[c_1TA+c_2+c_3\theta_{\tau'\sigma}]T[A_\tau-A] \;.
\end{split}
\ee
$c_1$ and $c_2$ are again determined from the initial conditions as~\eqref{c1c2fromInitial}. 
$c_3$ and $c_4$ are determined by demanding that $\phi_1$ and $\dot{\phi}_1$ be continuous at $\tau=\sigma$. With $\dot{t}(\sigma+\delta)=-\dot{t}(\sigma-\delta)$ and $\ddot{t}(\sigma+\delta)=\ddot{t}(\sigma-\delta)$ for $\delta\to0$ and $\delta>0$ we find
\be\label{c3c4BW}
\begin{split}
c_3&=-2(c_1TA_\sigma+c_2)+2\dot{t}_\LCp\ddot{t}_\sigma(c_1I_1+c_2I_0) \\
c_4&=-2(c_1I_1+c_2I_0) \;,
\end{split}
\ee
where $\dot{t}_\LCp:=\lim_{\delta\to0}\dot{t}(\sigma+\delta)$ and
\be
\begin{split}
I_0&=\int_0^{1/2}\frac{1}{\dot{t}^2}
\qquad
I_1=\int_0^{1/2}\frac{TA}{\dot{t}^2}
\\
I_2&=\int_0^{1/2}\frac{(TA)^2}{\dot{t}^2}=\frac{1}{2}-T^2m_\LCperp^2 I_0 \;.
\end{split}
\ee

We are assuming a symmetric field with a saddle point for the momentum integral at ${\bf p}={\bf k}/2$ and $k_3=0$, so 
\be
\dot{t}=\varepsilon_{\tau\sigma}T\sqrt{m_\LCperp^2+A^2(t)} \;,
\ee
$\sigma=1/2$, $t(1/2)=\tilde{t}$ where $A(\tilde{t})=i$, and $\dot{t}_\LCp=T\omega/2$.
This gives us all the information we need to obtain $\phi(1)$ and hence $\det\Lambda$. However, the expressions for $\phi(1)$ are rather long and complicated. This simplifies significantly, though, in the asymptotic limit, where we find
\be
\det\Lambda=-\frac{i\Omega A'(\tilde{t})}{2\pi_0(\infty)^3}t_0 \;.
\ee
Note that the limit $t_0\to\infty$ does not commute with $\Omega\to0$.

The instanton solution is given by
\be
\dot{x}^\LCperp=T(c^\LCperp+k^\LCperp\theta_{\tau\sigma}) 
\qquad
\dot{z}=T(A_3+c^3) \;,
\ee
where $c^j$ are three constants (which are completely unrelated to the constants in~\eqref{c1c2fromInitial} and~\eqref{c3c4BW}, which are only relevant for the calculation of $\det\Lambda$)
and
\be
\begin{split}
	&\tau<\sigma: \qquad \dot{t}=-T\sqrt{a_\LCm^2+c_\LCperp^2+(A_3+c^3)^2} \\
	&\tau>\sigma: \qquad \dot{t}=T\sqrt{a_\LCp^2+(c+k)_\LCperp^2+(A_3+c^3)^2} \;,
\end{split}
\ee
where $a_\LCp$ and $a_\LCm$ are two additional constants. $\dot{t}(\sigma+)-\dot{t}(\sigma-)=T\Omega$ gives the extra condition
\be
\begin{split}
&\sqrt{a_\LCm^2+c_\LCperp^2+(A_3(\tilde{t})+c^3)^2} \\
&+\sqrt{a_\LCp^2+(c+k)_\LCperp^2+(A_3(\tilde{t})+c^3)^2}=\Omega \;,
\end{split}
\ee
where $\tilde{t}=t(\sigma)$. We define
\be
\begin{split}
&G(a_\LCm^2,a_\LCp^2.c_j,\tilde{t}):=\int_{\tilde{t}}^{t_0}\ud t\sqrt{a_\LCm^2+c_\LCperp^2+(A_3+c^3)^2} \\
&+\int_{\tilde{t}}^{t_1}\ud t\sqrt{a_\LCp^2+(c+k)_\LCperp^2+(A_3+c^3)^2}+\Omega\tilde{t} \;.
\end{split}
\ee
The constants are determined from the remaining integration variables, $\sigma$, $T$ and $\Delta x^j=x^j(1)-x^j(0)$ by
\be\label{partialGconditionsPhoton}
\begin{split}
&\frac{\partial G}{\partial a_\LCm^2}=\frac{T\sigma}{2}
\qquad
\frac{\partial G}{\partial a_\LCp^2}=\frac{T(1-\sigma)}{2}
\\
&\frac{\partial G}{\partial\tilde{t}}=0
\qquad
\frac{\partial G}{\partial c_j}=-\Delta x^j \;.
\end{split}
\ee
The instanton action can we written as $e^{-iS}$ where (recall $k_3=0$)
\be\label{expa2cjPhoton}
\begin{split}
S=&\frac{T}{2}+\int_0^1\frac{\dot{x}^2}{2T}+A_3\dot{z}+Jx \\
=&\frac{T}{2}\left[\sigma(1-a_\LCm^2)+(1-\sigma)(1-a_\LCp^2)\right] \\
&-c_j\Delta x_j+G-x_\LCperp(1)k_\LCperp \;.
\end{split}
\ee

Now we perform the $\sigma$ integral with the saddle-point method. We have
\be
\frac{\ud S}{\ud\sigma}=\frac{T}{2}(a_\LCp^2-a_\LCm^2) \;,
\ee
where the terms with $\ud a_\LCpm^2/\ud\sigma$, $\ud\tilde{t}/\ud\sigma$ and $\ud c_j/\ud\sigma$ cancel due to~\eqref{partialGconditionsPhoton}. Thus, at the saddle point for $\sigma$ we have 
\be\label{amapsaddle}
a_\LCm^2(\sigma_{\rm saddle})=a_\LCp^2(\sigma_{\rm saddle}) \;, 
\ee
i.e. $\dot{x}^2$ is continuous at $\sigma$ even though $\dot{x}^\mu$ is not. To obtain the second derivative $\ud^2S/\ud\sigma^2$ we need $\ud a_\LCp^2/\ud\sigma$ and $\ud a_\LCm^2/\ud\sigma$. We can obtain these by differentiating~\eqref{partialGconditionsPhoton} and solving for $\ud a_\LCp^2/\ud\sigma$ and $\ud a_\LCm^2/\ud\sigma$ in terms of $(\partial/\partial c_\alpha)(\partial/\partial c_\beta)G$ with $\alpha,\beta=1,\dots,6$, where $c^4=\tilde{t}$, $c^5=a_\LCm^2$ and $c^6=a_\LCp^2$. However, at this intermediate stage, the result for $\ud^2S/\ud\sigma^2$ is quite complicated and not particularly illuminating. This is not a problem, because $\ud^2S/\ud\sigma^2$ goes into the pre-exponential, so even tough it depends on the remaining integration variables $T$, $\Delta x$ and ${\bf p}$, we actually only need it for the saddle point values $T\to T_{\rm saddle}$ etc. It also simplifies considerably when taking the asymptotic limit. We will therefore return to $\ud^2S/\ud\sigma^2$ once we have considered the $T$ and $\Delta x$ integrals.


Now the exponent becomes
\be
S=\frac{T}{2}(1-a^2)-c_j\Delta x_j+G-x_\LCperp(1)k_\LCperp \;,
\ee
where $a^2:=a_\LCm^2=a_\LCp^2$. Note that we do not actually need to find the explicit solution for the saddle point of $\sigma$. Instead of the first two equations in~\eqref{partialGconditionsPhoton} we have
\be\label{partialGconditionsPhotonVersT}
\frac{\partial G}{\partial a^2}=\frac{T}{2} \;.
\ee
We have
\be
\frac{\ud S}{\ud T}=\frac{1}{2}(1-a^2) \;,
\ee
where again all terms with $\ud a^2/\ud T$, $\ud c_j/\ud T$ and $\ud\tilde{t}/\ud T$ cancel. So, at the saddle point for $T$ we have 
\be\label{aTsaddle}
a^2(T_{\rm saddle})=1 \;,
\ee 
just as without the photon.
Again, the second derivative $\ud^2 S/\ud T^2$ can be calculated by differentiating the second line of equations in~\eqref{partialGconditionsPhoton} and~\eqref{partialGconditionsPhotonVersT} in order to solve for $\ud a^2/\ud T$ in terms of second-order partial derivatives of $G$. But we will again wait with the simplification of $\ud^2 S/\ud T^2$. 


Now the exponent is
\be\label{expWithMomentum}
\begin{split}
&ip'_j x^j(0)+ip_j x^j(1)-iS \\
&=i(p'+p-k)_jX^j+i(p'+c)_j\Delta x_j-iG \;,
\end{split}
\ee
where $X_j=(x(1)+x(0))_j/2$ and where we have used $\delta^3({\bf p}'+{\bf p}-{\bf k})$ to simplify the term proportional to $\Delta x_j$. We have
\be
\frac{\ud}{\ud\Delta x_j}\eqref{expWithMomentum}=i(p'+c)_j \;,
\ee
where the terms with $\ud c_k/\ud\Delta x_j$ cancel due to~\eqref{partialGconditionsPhoton}. Thus, at the saddle point for $\Delta x_j$ we have 
\be\label{cDxSaddle}
c_j(\Delta x_{\rm saddle})=-p'_j \;.
\ee
To obtain the pre-exponential we need 
\be\label{dcdD}
\frac{\ud}{\ud\Delta x_j}\frac{\ud}{\ud\Delta x_k}\eqref{expWithMomentum}=i\frac{\ud c_j}{\ud\Delta x_k} \;,
\ee 
which we will return to shortly.

The final exponent is given by
\be
\begin{split}
&i\int_{t_r}^{t_0}\pi_0(-{\bf p}')+i\int_{t_r}^{t_1}\pi_0({\bf p})-iG \\
&=i\int_{t_r}^{\tilde{t}}\ud t\left[\pi_0(-{\bf p}')+\pi_0({\bf p})-\Omega\right] \;, 
\end{split}
\ee
which is of course the same as what we have in~\eqref{finalExpGeneral}. 

We note again that we have obtained the final exponential without actually having to find $\sigma_{\rm saddle}$, $T_{\rm saddle}$ and $\Delta x_{\rm saddle}$. However, we will now turn to the pre-exponential contributions from these integrals and for this we need theses saddle points. So far we have used~\eqref{partialGconditionsPhoton} to determine the constants $a_\LCpm^2$ and $c^j$ and derivatives of these with respect to $\sigma$, $T$ and $\Delta x$. Although we do not yet have the saddle points, we know now that at these saddle points the constants are simply given by~\eqref{amapsaddle}, \eqref{aTsaddle} and~\eqref{cDxSaddle}. We can now obtain the saddle points by inserting these values of the constants into~\eqref{partialGconditionsPhoton}, e.g.
\be
T_{\rm saddle}=2\left(\frac{\partial G}{\partial a_\LCm^2}+\frac{\partial G}{\partial a_\LCp^2}\right)\bigg|_{a_\LCm^2=a_\LCp^2=1,{\bf c}=-{\bf p}'} \;.
\ee
Since we are now calculating the pre-exponential and since we will for simplicity consider the integrated probability rather than the momentum spectrum, we can simplify further by anticipating the saddle-point for the momentum integral, i.e. we set ${\bf p}={\bf k}/2$ (we have $A(-t)=-A(t)$). With $k_2=k_3=0$ we denote $p:=p_1=\Omega/2$. 
Even though we could calculate all the above quadratic terms for finite $t_0$, at the end we only need the asymptotic limit. For these terms we only need
\be
\begin{split}
G\to &t_0\sqrt{a_\LCp^2+c_\LCperp^2+(A_3(\infty)+c^3)^2} \\
+&t_1\sqrt{a_\LCm^2+(c+k)_\LCperp^2+(A_3(\infty)+c^3)^2} \;.
\end{split}
\ee
We choose for simplicity $t_0=t_1$.
We find
\be
\begin{split}
&T_{\rm saddle}\to\frac{2t_0}{\pi_0(\infty)}
\qquad
\sigma_{\rm saddle}\to\frac{1}{2}\\ 
&\Delta x^3_{\rm saddle}\to\frac{2t_0A(\infty)}{\pi_0(\infty)}
\qquad
\Delta x^\LCperp_{\rm saddle}\to0 \;,
\end{split}
\ee
where $\pi_0(\infty)=\sqrt{m_\LCperp^2+A^2(\infty)}$ and $m_\LCperp=\sqrt{1+p^2}$. By differentiating~\eqref{partialGconditionsPhoton} with respect to $\sigma$ and solving for $\ud a_\LCp^2/\ud\sigma$ and $\ud a_\LCm^2/\ud\sigma$ in terms of $(\partial/\partial c_\alpha)(\partial/\partial c_\beta)G$, we find
\be
\int\ud\delta\sigma\exp\left\{-\frac{4i(1+A^2(\infty))t_0}{\pi_0(\infty)}\delta\sigma^2\right\} \;,
\ee
where $\sigma=\sigma_{\rm saddle}+\delta\sigma$, and similarly
\be
\int\ud\delta T\exp\left\{-\frac{im_\LCperp^2\pi_0(\infty)}{4t_0}\delta T^2\right\} \;.
\ee
For the $\delta\Delta x$ integral we would in general have to calculate the determinant of~\eqref{dcdD}, but here it becomes diagonal,
\be
\begin{split}
\int\ud^3\delta\Delta x& \exp\bigg\{\frac{i\pi_0^3(\infty)}{4(1+A^2(\infty))t_0}\delta\Delta x_1^2\\
&+\frac{i\pi_0(\infty)}{4t_0}\delta\Delta x_2^2+\frac{i\pi_0^3(\infty)}{4m_\LCperp^2 t_0}\delta\Delta x_3^2\bigg\} \;.
\end{split}
\ee
The $(1+A_\infty^2)$ term might look unexpected, but it cancels\footnote{That contributions from different integrals cancel is not unexpected~\cite{Dunne:2006st,Ilderton:2015qda}. In fact, for Schwinger pair production in the closed-loop/probability-level instanton approach it has been shown that the different contributions combine into a single determinant similar to the Gutzwiller trace formula~\cite{Dietrich:2007vw}.} when collecting all the contributions from the Gaussian integrals, which gives (up to an irrelevant phase)
\be\label{intsCombBW}
\begin{split}
&\frac{1}{2}\frac{T}{(2\pi T)^2}\frac{1}{\sqrt{\det\Lambda}}\int\ud\sigma\int\ud T\int\ud^3\Delta x^j \\
&\to\sqrt{\frac{\pi}{2A'(\tilde{t})\Omega}}\frac{1}{\pi_0(\infty)} \;.
\end{split}
\ee
We have included an extra factor of $T$ here compared to the no-photon case, because such a factor comes from $-i\epsilon^{(\LCpara)}\dot{q}\to T$ ($-i\epsilon^{(\LCperp)}\dot{q}\to 0$) and from the term linear in $\epsilon$ form the spin factor $\mathcal{P}\exp(\dots)$. As we already noted, the limits $t_0\to\infty$ and $\Omega\to0$ do not commute. In particular, $\det\Lambda$ has a different $t_0$ scaling for $\omega>0$ and $\omega=0$. This different scaling is needed in order to cancel the extra factors of $t_0$ coming from $\int\ud\sigma$ and the extra overall factor of $T$ to give a finite limit as $t_0\to\infty$.

For the prefactor we also need
\be\label{epsScalarPartBW}
\begin{split}
&\mathcal{P}\exp\left\{-\frac{iT}{4}\int_0^1\sigma^{\mu\nu}F_{\mu\nu}\right\}=\exp\left\{\frac{T}{2}\int_0^1\ud\tau A_3'(t)\gamma^0\gamma^3\right\} \\
&=\exp\left\{-\ln\left[\frac{i+p}{A_\infty+\pi_0(\infty)}\right]\gamma^0\gamma^3\right\} \;,
\end{split}
\ee
and
\be\label{epsSpinorPartBW}
\begin{split}
&\mathcal{P}\exp\left\{\int_0^1\frac{TA'}{2}\gamma^0\gamma^3-\frac{iT}{2}\slashed{k}\slashed{\epsilon}e^{-ikq}\right\} \\
&\to\int_0^1\ud\sigma\exp\left\{\int_\sigma^1\frac{TA'}{2}\gamma^0\gamma^3\right\}\left(-\frac{iT}{2}\slashed{k}\slashed{\epsilon}e^{-ikq}\right) \\
&\hspace{1.5cm}\times\exp\left\{\int_0^\sigma\frac{TA'}{2}\gamma^0\gamma^3\right\} \;,
\end{split}
\ee
where
\be
\int_0^\sigma\frac{TA'}{2}=\int_\sigma^1\frac{TA'}{2}=-\frac{1}{2}\ln\left[\frac{i+p}{A_\infty+\pi_0(\infty)}\right] \;.
\ee

\subsection{Momentum integrals and final results}

We finally find
\be
\begin{split}
M=&(2\pi)^3\delta^3({\bf p}+{\bf p}'-{\bf k})\mathcal{A} \\
\times&\exp\left\{i\int_{t_r}^{\tilde{t}}\ud t\left[\pi_0(-{\bf p}')+\pi_0({\bf p})-\Omega\right]\right\} \;,
\end{split}
\ee
where $\mathcal{A}$ is a pre-exponential factor which depends on the photon polarization. For a photon with momentum $k_\mu=\Omega\{1,\sin\theta,0,\cos\theta\}$ we can choose $\epsilon^{(\LCpara)}_\mu=\{0,-\cos\theta,0,\sin\theta\}$ and $\epsilon^{(\LCperp)}_\mu=\{0,0,1,0\}$ as basis for the polarization vector. A general polarization vector can then be expressed as
\be\label{epsilonrholambda}
\epsilon_\mu=\cos\left[\frac{\rho}{2}\right]\epsilon^{(\LCpara)}_\mu+\sin\left[\frac{\rho}{2}\right]e^{i\lambda}\epsilon^{(\LCperp)}_\mu \;,
\ee
where $\rho$ and $\lambda$ are two real constants. The reason for choosing $\rho/2$ in~\eqref{epsilonrholambda} is because then the polarization dependence on the probability level can be expressed in terms of the following Stokes vector (cf.~\cite{Dinu:2018efz})
\be\label{stokesDefinition}
{\bf N}=\{1,\cos\lambda\sin\rho,\sin\lambda\sin\rho,\cos\rho\} \;.
\ee
Summing over the fermion spins we find
\be\label{ampSquaredSummed}
\sum_\text{spins}|\mathcal{A}|^2=\frac{2\pi}{\Omega A'(\tilde{t})(1+p^2)}\{1+3p^2,0,0,1-p^2\}\cdot{\bf N} \;,
\ee
where $p=\Omega/2$. We see that the smallest and largest probabilities are obtained for parallel and perpendicular linear polarization, ${\bf N}=\{1,0,0,\pm1\}$, while e.g. both left- and right-handed circular polarization, ${\bf N}=\{1,0,\pm1,0\}$, gives the same probability.

Note that, while the exponent contains the full momentum dependence, we have only calculated the prefactor at the saddle point ${\bf p}={\bf p}'={\bf k}/2$ (and we have assumed $k_3=0$). It is straightforward to check that~\eqref{ampSquaredSummed} agrees with what one finds with the WKB approach. To obtain the full prefactor we have to perform the momentum integrals, but we can already see that the ratio perpendicular/parallel is independent of the field. We have assumed $A'(\pm\infty)=0$ here, but we have the same ratio for a constant field.

We obtain the total prefactor by expanding the exponent around the saddle point ${\bf p}={\bf k}/2$. We change variable from ${\bf p}=\frac{\bf k}{2}+\frac{\bm\Delta}{2}$ to ${\bm\Delta}$. The term that is linear in $\Delta_3$ vanishes because $\tilde{t}$ is imaginary and so
\be
\text{Re} i\int_0^{\tilde{t}}\ud t\frac{A}{\sqrt{m^2+A^2}}=0 \;.
\ee  
We find that the probability can be expressed as
\be
P={\bf M}\cdot{\bf N} 
\ee
with
\be\label{finalGeneralPhotonTimeE}
\begin{split}
{\bf M}=&\frac{\sqrt{\pi}\alpha \omega^{3/2}}{4\sqrt{2}\Omega^2m_\LCperp^2 A'(\tilde{t})}\{1+3p^2,0,0,1-p^2\}\\
&\times\frac{\exp\left\{-\frac{4}{\omega}(\mathcal{J}_0-p\tilde{u})\right\}}{\sqrt{\mathcal{J}_1(\mathcal{J}_1-p^2\mathcal{J}_2)\left(m_\LCperp^2\mathcal{J}_2-\frac{\omega}{p A'(\tilde{t})}\right)}} \;,
\end{split}
\ee
where $A(\tilde{t})=i$ and
\be
\mathcal{J}_n=\int_0^{\tilde{u}}\ud u\left(m_\LCperp^2-\frac{\tilde{f}^2(u)}{\gamma^2}\right)^{\frac{1}{2}-n} \;,
\ee
where $A(t=iu/\omega)=:i\tilde{f}(u)/\gamma$ ($\gamma=\omega/E$) and $\tilde{f}(\tilde{u})=\gamma$. For example, for a Sauter pulse $A(t)=\frac{1}{\gamma}\tanh(\omega t)$ and $\tilde{f}(u)=\tan u$. 

To obtain all terms, including e.g. the one with $\gamma E/(p A')$, we expanded
\be
\begin{split}
\tilde{t}({\bf p})=&\frac{1}{\omega}f^{-1}\left(\gamma\left[i\sqrt{1+\left(\frac{\Delta_2}{2}\right)^2}+\frac{\Delta_3}{2}\right]\right) \\
=&\tilde{t}_0+\delta\tilde{t}_1+\delta^2\tilde{t}_2 \;,
\end{split}
\ee
where $A(t)=:f(\omega t)/\gamma$ and $\delta=\mathcal{O}(\Delta_j)$ is a bookkeeping parameter, used
\be
\int_0^{\tilde{t}}\ud t F(t)\approx\int_0^{\tilde{t}_0}\ud t F(t)+F(\tilde{t}_0) (\delta\tilde{t}_1+\delta^2\tilde{t}_2)+\frac{F'(t_0)}{2}\delta^2\tilde{t}_1^2
\ee
and then expanded the integrand $F$ to second order in $\Delta_j\sim\delta$.

Since $\tilde{u}$ does not depend on $p$ (in the final result~\eqref{finalGeneralPhotonTimeE}), we have
\be\label{Jfromdp}
\mathcal{J}_1=\frac{1}{p}\frac{\ud\mathcal{J}_0}{\ud p} 
\qquad
\mathcal{J}_2=-\frac{1}{p}\frac{\ud\mathcal{J}_1}{\ud p} \;,
\ee
so if we have chosen a field for which we can obtain $\mathcal{J}_0$ analytically, then that is the only integral we need to perform. We can also obtain $\mathcal{J}_1$ and $\mathcal{J}_2$ by instead differentiating with respect to $\gamma$,
\be\label{Jfromdgamma}
\begin{split}
\mathcal{J}_1&=\frac{1}{m_\LCperp^2}\left(\frac{\ud}{\ud\gamma}[\gamma\mathcal{J}_0]-\frac{\omega p}{A'(\tilde{t})}\right) \\
\mathcal{J}_2&=\frac{1}{m_\LCperp^2}\left(\frac{\omega}{pA'(\tilde{t})}-\gamma^2\frac{\ud}{\ud\gamma}\frac{\mathcal{J}_1}{\gamma}\right) \;.
\end{split}
\ee

It is straightforward to check that for a Sauter pulse we find 
\begin{widetext}
\be\label{SauterPreFin}
{\bf M}_{\rm Sauter}=\frac{\alpha\sqrt{\pi E}(1+m_\LCperp^2\gamma^2)^{7/4}\exp\left\{-\frac{4}{E\gamma^2}\left(\sqrt{1+m_\LCperp^2\gamma^2}\Lambda-\text{arccot}(p)-\gamma p\text{arctan}(\gamma)\right)\right\}}{8\Omega p m_\LCperp\gamma\sqrt{1+\gamma^2}\sqrt{2\Lambda}\left(m_\LCperp^2(1+\gamma^2)\Lambda-p\sqrt{1+m_\LCperp^2\gamma^2}\right)}\{1+3p^2,0,0,1-p^2\} \;,
\ee
\end{widetext}
where
\be
\Lambda=\text{arctan}\left[\frac{1}{p}\sqrt{1+m_\LCperp^2\gamma^2}\right] \;.
\ee

In the limit of a slowly varying field we have for a general pulse with a maximum at $t=0$ ($E'(0)=0$)
\be\label{MBWslow}
\begin{split}
\lim_{\gamma\ll1}{\bf M}=&\frac{\sqrt{\pi}\alpha E^2(0)\{1+3p^2,0,0,1-p^2\}}{8\sqrt{-E''(0)}\Omega p m_\LCperp} \\
\times&\frac{\exp\left\{-\frac{2}{E(0)}\left(m_\LCperp^2\text{arccot}(p)-p\right)\right\}}{\sqrt{\text{arccot}(p)}[m_\LCperp^2\text{arccot}(p)-p]}
\end{split}
\ee
We can obtain this from the constant field result~\eqref{PparaperpConstant} by replacing the volume factor with a time integral, $V_0\to\int dt$, the constant field strength with a locally constant one, $E\to E(t)$, and then performing the time integral with the saddle-point method, i.e. expanding $\frac{1}{E(t)}\approx\frac{1}{E(0)}-\frac{E''(0)}{2E'(0)}t^2$ in the exponential part of the integrand. Thus, the usual LCF ideas work also in this case.

In the low-energy limit we can use~\eqref{Jfromdgamma} to obtain
\be\label{MBWlow}
\lim_{\Omega\ll1}{\bf M}=\frac{\alpha E^{5/2}\{1,0,0,1\}\exp\left\{-\frac{\pi}{E}g\right\}}{2\sqrt{2}\pi\Omega^2A'(\tilde{t})\omega\partial_{\gamma^2}(\gamma^2g)\sqrt{-\partial^2_{\gamma^2}(\gamma^2g)}} \;,
\ee
where 
\be
\begin{split}
g(\gamma)&=\frac{4}{\pi\gamma}\int_0^{\tilde{u}}\ud u\left(1-\frac{\tilde{f}^2(u)}{\gamma^2}\right)^{\frac{1}{2}} \\
&=\frac{4}{\pi}\int_0^1\ud y\frac{\sqrt{1-y^2}}{\tilde{f}'} \;,
\end{split}
\ee
where in the second line we have changed integration variable to $y=\tilde{f}(u)/\gamma$. This is exactly the same $g$ as in~\cite{Dunne:2006st} for Schwinger pair production. In fact, $g$ enters~\eqref{MBWlow} in exactly the same way as Eq.~(3.58) in~\cite{Dunne:2006st}, i.e. both in the exponential and the pre-exponential factors. This is related to the fact that an electric field can produce a pair without the photon, and many of the integrals will be the same in the $\Omega\to0$ limit as in the complete absence of this photon. There is though an additional field dependence in~\eqref{MBWlow} due to $A'(\tilde{t})$ in the pre-exponential. For a Sauter pulse, we have
\be
\begin{split}
\lim_{\Omega\ll1}{\bf M}_{\rm Sauter}=&\frac{\alpha\sqrt{E}(1+\gamma^2)^{1/4}}{2\pi\Omega^2\gamma}\{1,0,0,1\} \\
\times&\exp\left\{-\frac{\pi}{E}\frac{2}{1+\sqrt{1+\gamma^2}}\right\} \;,
\end{split}
\ee
which can be obtained either by taking the $\omega\to0$ limit of~\eqref{SauterPreFin} or by evaluating~\eqref{MBWlow} for a Sauter pulse.

In the high-energy limit we find
\be\label{MBWplaneWave}
\lim_{\Omega\gg1}{\bf M}=\frac{\alpha\sqrt{\pi\chi}\{3,0,0,-1\}\exp\left\{-\frac{4a_0}{\chi}(\tilde{u}-a_0^2\mathcal{J}_{\rm PW})\right\}}{32a_0\sqrt{\tilde{u}\tilde{f}'(\tilde{u})(\tilde{u}-a_0^2\mathcal{J}_{\rm PW})(a_0\tilde{u}\tilde{f}'(\tilde{u})-1)}} \;,
\ee
where
\be
\mathcal{J}_{\rm PW}=\int_0^{\tilde{u}}\ud u\; \tilde{f}^2(u) \;.
\ee
This agrees exactly with the result for a plane-wave background field, which can be obtained as follows: The probability for nonlinear Breit-Wheeler pair production in an arbitrary plane wave and for arbitrary parameters can be obtained from Eq.~(35), (36) and (39) in~\cite{Dinu:2019pau}. Those expressions contain integrals over two lightfront-time variables, $\phi=(\phi_2+\phi_1)/2$ and $\theta=\phi_2-\phi_1$, and over one longitudinal momentum $s_2$. For $\chi<1$ these integrals can be performed with the saddle-point method, with a saddle point at $s_2=1/2$ (the electron and positron share the initial longitudinal momentum equally), $\phi=0$ (average lightfront time at field maximum) and $\theta=2i\tilde{u}$.

For a Sauter pulse we find
\be
\begin{split}
\lim_{\Omega\gg1}{\bf M}_{\rm Sauter}=&\frac{\alpha\sqrt{\pi a_0\chi}\{3,0,0,-1\}}{32\sqrt{(1+a_0^2)\text{arccot}(a_0)}} \\
\times&\frac{\exp\left\{-\frac{4a_0}{\chi}[(1+a_0^2)\text{arccot}(a_0)-a_0]\right\}}{(1+a_0^2)\text{arccot}(a_0)-a_0} \;,
\end{split}
\ee
which can be obtained either by evaluating~\eqref{MBWplaneWave} for a Sauter pulse, or by taking the high-energy limit of~\eqref{SauterPreFin}.

Perhaps the experimentally most relevant limit is a slow field and high energy,
\be\label{MBWPWlcf}
\lim_{\gamma\ll1}\lim_{\Omega\gg1}{\bf M}=\frac{3\alpha\sqrt{\pi\Omega}E^2(0)}{32\sqrt{-2E''(0)}}\{3,0,0,-1\}e^{-\frac{8}{3\Omega E(0)}} \;,
\ee
which can be obtained either from the locally-constant-field ($a_0\gg1$) limit of the high-energy/plane-wave approximation~\eqref{MBWplaneWave} or from the high-energy limit of the locally-constant-electric-field approximation~\eqref{MBWslow}.

From~\eqref{MBWlow} we see that for low energy $\Omega\ll1$ the probability is maximized by parallel polarization, ${\bf N}=\{1,0,0,1\}$, but vanishes for perpendicular polarization, ${\bf N}=\{1,0,0,-1\}$. In contrast, in the high-energy limit~\eqref{MBWplaneWave} the probability for perpendicular polarization is twice as large, $(3+1)/(3-1)=2$. This is known in the LCF limit of plane waves~\eqref{MBWPWlcf}, see~\cite{Reiss62,Nikishov:1964zza,Ritus1985}, but we see from~\eqref{MBWplaneWave} that this holds in general in the high-energy limit.

\section{Prefactor for nonlinear Compton scattering in time-dependent electric fields}\label{Prefactor for nonlinear Compton}

The calculation for nonlinear Compton is very similar. Before we begin we mention that nonlinear Compton in a time-dependent electric field has recently been studied in~\cite{GelferPresentation}, where a WKB approach was used and where the exponential part of the time integrand was expanded to cubic order to obtain results in terms of Airy functions and in particular to compare and check the LCF approximation. For exact expressions for nonlinear Compton in an electric field see~\cite{NikishovReview1985}. 

\subsection{Worldline derivation}

We assume the initial electron travels perpendicular to the field, with $p:=p_1>0$ and $p_2=p_3=0$. Instead of~\eqref{LSZ3pair} we have
\be\label{LSZ3Compton}
\begin{split}
M=\lim_{t\to-\infty}\lim_{t'\to+\infty}&\int\ud^3x\ud^3 x'\,e^{ip'_jx'^j}\bar{u}_r^{(\rm asymp)}(t',{\bf p}')\gamma^0\\
\times&S(x',x)\gamma^0 e^{-ip_jx^j}u_{r}^{(\rm asymp)}(t,{\bf p}) \;,
\end{split}
\ee
where the Compton amplitude is obtained by replacing $A\to A+\epsilon e^{-ikx}$ and selecting the term that is linear in $\epsilon_\mu$. So, for the exponent we can obtain most results from the pair-production case by replacing ${\bf p}'\to-{\bf p}$, ${\bf p}\to{\bf p}'$ and $k_\mu\to-k_\mu$. Instead of~\eqref{dotq0sol} we have
\be\label{dotq0solCompton}
\begin{split}
\dot{q}_0(\tau)=&\theta_{\sigma\tau}T\sqrt{1+({\bf p}-{\bf A}(q_0(\tau)))^2}\\
&+\theta_{\tau\sigma}T\sqrt{1+({\bf p}'-{\bf A}(q_0(\tau)))^2} 
\end{split}
\ee
and instead of~\eqref{eqfortildet} we have
\be\label{eqfortildetCompton}
\sqrt{1+({\bf p}'-{\bf A}(\tilde{t}))^2}\overset{!}{=}\sqrt{1+({\bf p}+{\bf A}(\tilde{t}))^2}-\Omega \;.
\ee

For the prefactor we again have~\eqref{phi1BW}, \eqref{phi2BW} and~\eqref{c1c2fromInitial}. To calculate the rest of the prefactor we put $k_2$ and $k_3$ (or equivalently, due to momentum conservation, $p'_{2,3}$) equal to their saddle point value, which, as we will show, is $k_2=k_3=0$. We keep the general dependence on the component that is parallel to the initial electron's momentum, i.e. $k_1$. One reason for not integrating over $k_1$ is that that would in general lead to an IR divergence and, even if IR finite, the result would receive dominant contribution from soft photons and would therefore not have an exponential scaling. In other words, the saddle-point method only works if we prevent the photon from being too soft. Exponential approximations for the emission of hard photons have been considered in~\cite{Dinu:2018efz,Blackburn:2017dpn,HernandezAcosta:2020agu}. We will focus on the spectrum, but obtaining the probability integrated over $k_1$ with some lower cut off, $|k_1|>k_c$, is straightforward to obtain by simply expanding the integrand around this cut off. In particular, the exponential part is the same, i.e. if $P(k_1)\sim\exp(-f(k_1)/E)$ then $\int_{k_c}^\infty\ud k_1 P(k_1)\sim\exp(-f(k_c)/E)$. 

At $k_2=k_3=0$, \eqref{eqfortildetCompton} reduces to $A(\tilde{t})=i$,
so we have the same $\tilde{t}$ as in the pair-production case. We assume $0<p-\Omega<p$.

We once again perform the path integral with the Gelfand-Yaglom method with~\eqref{pathQuadLambda}, \eqref{LambdaEq} and~\eqref{phiAnsatz}. The calculation is very similar to the one in the previous section for Breit-Wheeler. The solution of~\eqref{LambdaEq} is again given by~\eqref{phi1BW} and~\eqref{phi2BW} with $c_1$ and $c_2$ given by~\eqref{c1c2fromInitial}. We have an implicit difference due to the fact that the instanton $t(\tau)$ is different. We also have an explicit difference in the form of $c_3$ and $c_4$. These two constants are again determined by demanding that $\phi(\tau)$ and $\dot{\phi}(\tau)$ be continuous at $\tau=\sigma$. For this we need
\be
\dot{t}(\sigma-)=Tp \qquad \dot{t}(\sigma+)=T(p-\Omega) 
\ee 
and $\ddot{t}(\sigma-)=\ddot{t}(\sigma+)=iT^2A'(\tilde{t})$, which follows from~\eqref{dotq0solCompton} and $A(\tilde{t})=i$. We find
\be
\begin{split}
c_3&=-\Omega\left(\frac{iTc_1+c_2}{p}+iT^3A'(\tilde{t})\int_0^\sigma\frac{\ud\tau}{\dot{t}^2}(c_1TA+c_2)\right)\\
c_4&=\frac{\Omega}{p-\Omega}\int_0^\sigma\frac{\ud\tau}{\dot{t}^2}(c_1TA+c_2)  \;.
\end{split}
\ee
The determinant~\eqref{detLambdaIstart} can now be expressed in terms of
\be
I_n=\int_0^\sigma\ud\tau\frac{(TA)^n}{\dot{t}^2}=\int_{t_0}^{\tilde{t}}\ud t\frac{(TA)^n}{T^3\sqrt{m_\LCperp^2+A^2}}
\ee 
and
\be
J_n=\int_\sigma^1\ud\tau\frac{(TA)^n}{\dot{t}^2}=\int_{\tilde{t}}^{t_1}\ud t\frac{(TA)^n}{T^3\sqrt{m_\LCperp^{\prime2}+A^2}} \;.
\ee

While the intermediate steps involve some rather long expressions, things simplify considerably in the asymptotic limit.
We have
\be
\sigma T=T\int_0^\sigma\ud\tau=\int_{t_0}^{\tilde{t}}\frac{\ud t}{\sqrt{m^2+A^2}}\to\frac{-t_0}{\pi_0(\infty)}
\ee
and
\be
(1-\sigma)T=T\int_\sigma^1\ud\tau=\int_{\tilde{t}}^{t_1}\frac{\ud t}{\sqrt{m^{\prime2}+A^2}}\to\frac{t_1}{\pi'_0(\infty)} \;,
\ee
so
\be\label{TasymptoticCompton}
T\to\frac{-t_0}{\pi_0(\infty)}+\frac{t_1}{\pi'_0(\infty)}
\ee
and
\be\label{sigmaasymptoticCompton}
\sigma\to\frac{-t_0}{\pi_0(\infty)}\left(\frac{-t_0}{\pi_0(\infty)}+\frac{t_1}{\pi'_0(\infty)}\right)^{-1} \;,
\ee
where (we have assumed $A(-t)=-A(t)$) 
\be
\begin{split}
\pi_0(\infty)&=\sqrt{1+p^2+A^2(\infty)}\\ 
\pi'_0(\infty)&=\sqrt{1+(p-\Omega)^2+A^2(\infty)} \;.
\end{split}
\ee
Similarly, 
\be
I_n\to\frac{-t_0(TA)^n}{T^3\pi_0^3(\infty)}
\qquad
J_n\to\frac{t_1(TA)^n}{T^3\pi_0^{\prime3}(\infty)} \;.
\ee
If we let $L$ be a large parameter such that $-t_0=\mathcal{O}(L)$ and $t_1=\mathcal{O}(L)$, then we can expand~\eqref{detLambdaIstart} by taking into account $\dot{t}(\tau=0,1)=\mathcal{O}(L)$, $T=\mathcal{O}(L)$, $I_n=\mathcal{O}(L^{n-2})$ and $J_n=\mathcal{O}(L^{n-2})$. 
We find
\be
\det\Lambda=\frac{i\Omega A'(\tilde{t}) t_0t_1}{T\pi_0^2\pi_0^{\prime2}} \;,
\ee
We note again that the asymptotic limit does not commute with $\Omega\to0$.

When calculating $\det\Lambda$ we could replace the remaining integration variables with their saddle-point values since $\det\Lambda$ only contributes to the pre-exponential factor. Now we turn to the remaining integrals. Instead of~\eqref{dotq0solCompton} we need the instanton solution before replacing $T\to T_{\rm saddle}$ etc., which is given by
\be
\dot{x}^\LCperp=T(c^\LCperp-k^\LCperp\theta_{\tau\sigma}) 
\qquad
\dot{z}=T(A_3+c^3-k^3\theta_{\tau\sigma})
\ee
and
\be
\begin{split}
	&\tau<\sigma: \qquad \dot{t}=T\sqrt{a_\LCm^2+c_\LCperp^2+(A_3+c^3)^2} \\
	&\tau>\sigma: \qquad \dot{t}=T\sqrt{a_\LCp^2+(c-k)_\LCperp^2+(A_3+c^3-k^3)^2} \;,
\end{split}
\ee
where $\dot{t}(\sigma+)-\dot{t}(\sigma-)=-T\Omega$ gives the extra condition
\be
\begin{split}
&\sqrt{a_\LCm^2+c_\LCperp^2+(A_3(\tilde{t})+c^3)^2} \\
=&\sqrt{a_\LCp^2+(c-k)_\LCperp^2+(A_3(\tilde{t})+c^3-k^3)^2}+\Omega \;,
\end{split}
\ee
where $\tilde{t}=t(\sigma)$. We define
\be
\begin{split}
&G(a_\LCm^2,a_\LCp^2.c_j,\tilde{t}):=\int_{t_0}^{\tilde{t}}\ud t\sqrt{a_\LCm^2+c_\LCperp^2+(A_3+c^3)^2} \\
&+\int_{\tilde{t}}^{t_1}\ud t\sqrt{a_\LCp^2+(c-k)_\LCperp^2+(A_3+c^3-k^3)^2}-\Omega\tilde{t} \;.
\end{split}
\ee
The constants are again determined from $\sigma$, $T$ and $\Delta x^j$ by the same equations as in~\eqref{partialGconditionsPhoton}. We have 
\be\label{expa2cjPhotonCompton}
\begin{split}
S=&\frac{T}{2}+\int_0^1\frac{\dot{x}^2}{2T}+A_3\dot{z}+Jx \\
=&\frac{T}{2}\left[\sigma(1-a_\LCm^2)+(1-\sigma)(1-a_\LCp^2)\right] \\
&-c_j\Delta x_j+G+x_j(1)k_j \;.
\end{split}
\ee

This exponent has the same form as in the Breit-Wheeler case, and so the $\sigma$ and $T$ integrals are performed in exactly the same way as before, i.e. by differentiating~\eqref{partialGconditionsPhoton} with respect to the integration variables and solving for e.g. $\ud a^2/\ud T$. 

Instead of~\eqref{expWithMomentum} we have 
\be\label{expWithMomentumCompton}
\begin{split}
&-ip_j x^j(0)+ip'_j x^j(1)-iS \\
&=i(-p+p'+k)_jX^j+i(c-p)_j\Delta x_j-iG \;,
\end{split}
\ee
where $X_j=(x(1)+x(0))_j/2$ and where we have used $\delta^3({\bf p}'+{\bf k}-{\bf p})$ to simplify the term proportional to $\Delta x_j$. We have
\be
\frac{\ud}{\ud\Delta x_j}\eqref{expWithMomentumCompton}=i(c-p)_j \;,
\ee
where the terms with $\ud c_k/\ud\Delta x_j$ cancel. Thus 
\be
c_j(\Delta x_{\rm saddle})=p_j
\ee
at the saddle point for $\Delta x_j$. 

The final exponent for Compton scattering is given by
\be
\begin{split}
&-i\int_{t_r}^{t_0}\pi_0({\bf p})+i\int_{t_r}^{t_1}\pi_0({\bf p}')-iG \\
&=i\int_{t_r}^{\tilde{t}}\ud t\left[-\pi_0({\bf p})+\pi_0({\bf p}')+\Omega\right] \;,
\end{split}
\ee
which agrees, as it should, with what one finds with the WKB approach.

To perform the $\Delta x$ integral we need the second derivatives, given by
\be\label{dcdDCompton}
\frac{\ud}{\ud\Delta x_j}\frac{\ud}{\ud\Delta x_k}\eqref{expWithMomentum}=i\frac{\ud c_j}{\ud\Delta x_k} \;,
\ee
which can be obtained by differentiating~\eqref{partialGconditionsPhoton}. The $\Delta x_j$ integral gives $(2\pi)^{3/2}/\sqrt{\det(\eqref{dcdDCompton})}$. The intermediate results for the $\sigma$, $T$ and $\Delta x$ integrals are even more complicated for Compton scattering compared the results in Breit-Wheeler, which were already rather complicated. However, this again simplifies considerably in the asymptotic limit and when replacing all the integration variables with their saddle points. We can obtain these from~\eqref{partialGconditionsPhoton} since we know that at these saddle points we have $a_\LCm^2=a_\LCp^2=1$ and ${\bf c}={\bf p}$. We have also assumed $p_2=p_3=0$ and we set $p_1=p$. For the pre-exponential factor we can also set $k_2=k_3=0$ and $k_1=\Omega$. To obtain the asymptotic limit we can use
\be
\begin{split}
G\to&-t_0\sqrt{a_\LCm^2+c_\LCperp^2+(c^3-A)^2}\\
&+t_1\sqrt{a_\LCp^2+(c-k)_\LCperp^2+(c^3-k^3+A)^2} \;,
\end{split}
\ee
where $A=A_3(\infty)$ (recall that we have assumed $A(-t)=-A(t)$).
From~\eqref{partialGconditionsPhoton} we find $T_{\rm saddle}$ and $\sigma_{\rm saddle}$ as in~\eqref{TasymptoticCompton} and~\eqref{sigmaasymptoticCompton}, and
\be\label{Deltax1}
\Delta x_1\to\frac{-t_0}{\pi_0(\infty)}p+\frac{t_1}{\pi'_0(\infty)}p' \;,
\ee
where $p'=p-\Omega$, $\Delta x^2\to0$ and
\be
\Delta x_3\to-\left(\frac{t_0}{\pi_0(\infty)}+\frac{t_1}{\pi'_0(\infty)}\right)A(\infty) \;.
\ee
While the dynamics at finite times gives a nontrivial contribution to $\Delta x$, for $-t_0,t_1\to\infty$ the dominant contribution comes just from the asymptotic parts of the instanton, where it is outside the electric field $A'(t)$. So, \eqref{Deltax1} is the distance one should expect for a particle that initially has momentum $p$ for a proper-time interval of length $\Delta\tau\sim-t_0/\pi_0(\infty)$ and then momentum $p'$ for the second half with $\Delta\tau\sim t_1/\pi'_0(\infty)$.

Collecting the contributions from all the integrals we find
\be\label{intsCombC}
\begin{split}
&\frac{1}{2}\frac{T}{(2\pi T)^2}\frac{1}{\sqrt{\det\Lambda}}\int\ud\sigma\int\ud T\int\ud^3\Delta x^j \\
&\to\sqrt{\frac{\pi}{2A'(\tilde{t})\Omega\pi_0(\infty)\pi'_0(\infty)}} \;.
\end{split}
\ee
This is very similar to the Breit-Wheeler case, as we obtain~\eqref{intsCombBW} by simply replacing $\pi'_0\to\pi_0$ in~\eqref{intsCombC} (note that in the Breit-Wheeler case we calculated the prefactor at the saddle point where ${\bf p}={\bf p}'$).

To calculate the spin part of the pre-exponential as in~\eqref{epsScalarPartBW} and~\eqref{epsSpinorPartBW} we need
\be
\int_0^\sigma\frac{TA'}{2}=\frac{1}{2}\ln\left[\frac{i+p}{-A(\infty)+\pi_0(\infty)}\right]
\ee
and
\be
\int_\sigma^1\frac{TA'}{2}=-\frac{1}{2}\ln\left[\frac{i+p'}{A(\infty)+\pi'_0(\infty)}\right] \;.
\ee

\subsection{Results}

We finally find
\be
\begin{split}
M=&(2\pi)^3\delta^3({\bf p}'+{\bf k}-{\bf p})\mathcal{A} \\
\times&\exp\left\{i\int_{t_r}^{\tilde{t}}\ud t\left[-\pi_0({\bf p})+\pi_0({\bf p}')+\Omega\right]\right\}
\end{split}
\ee
where averaging over the initial spin and summing over the final spin gives
\be\label{ampSquaredSummedCompton}
\begin{split}
&\frac{1}{2}\sum_\text{spins}|\mathcal{A}|^2=\frac{\pi}{2m_\LCperp m'_\LCperp\Omega A'(\tilde{t})}\\
&\times\left\{m_\LCperp^2+{m'}_\LCperp^2+\Omega^2,0,0,m_\LCperp^2+{m'}_\LCperp^2-\Omega^2\right\}\cdot{\bf N} \;,
\end{split}
\ee
where ${\bf N}$ is the Stokes vector for the photon polarization in~\eqref{stokesDefinition}. From here on the calculations are the same as in a WKB approach.

To avoid IR/soft photon contributions, we will keep one component of the photon momentum fixed and integrate over the other two. One option would be to keep $k_1$ fixed and integrate over $k_2$ and $k_3$, with a saddle point at $k_2=k_3=0$. However, noting that $\chi_\gamma:=\sqrt{-(Fk)^2}=E(k_1^2+k_2^2)$, we will instead change to cylindrical coordinates, $k_1={\sf k}\cos\varphi$ and $k_2={\sf k}\sin\varphi$, and perform the $k_3$ and $\varphi$ integrals with the saddle-point method, with a saddle point at $k_3=\varphi=0$. 
We define the longitudinal momentum spectrum as the integrand in the total probability
\be
P=\int\ud {\sf k} P({\sf k}) \;.
\ee
Since we do not integrate over one momentum variable, the final results will depend on one more parameter compared to the Breit-Wheeler results in the previous section, which therefore leads to more complicated expressions. So, we consider for simplicity a Sauter pulse. 
The polarization dependence can be expressed in terms of a Stokes vector as $P({\sf k})={\bf N}\cdot{\bf M}(\sf k)$, where 
\begin{widetext}
\be\label{preSauterCompton}
\begin{split}
{\bf M}_{\rm Sauter}({\sf k})=&\frac{\alpha(1+(\gamma m'_\LCperp)^2))^{1/4}\{m_\LCperp^2+m_\LCperp^{\prime2}+{\sf k}^2,0,0,m_\LCperp^2+m_\LCperp^{\prime2}-{\sf k}^2\}\exp[f(p')-f(p)]}{8m_\LCperp m'_\LCperp{\sf k}(1+\gamma^2)\sqrt{p\text{ arctan}\left[\frac{\sqrt{1+(\gamma m'_\LCperp)^2}}{p'}\right]}}\\
\times&\left(\frac{\text{arctan }\gamma}{\gamma}+{\sf k}(\gamma m'_\LCperp)^2\frac{\text{arctan}\left[\frac{\sqrt{1+(\gamma m'_\LCperp)^2}}{p'}\right]}{(1+(\gamma m'_\LCperp)^2)^{3/2}}+\frac{1}{p'}\left[\frac{{\sf k}}{1+(\gamma m'_\LCperp)^2}-\frac{p}{1+\gamma^2}\right]\right)^{-1/2}
\end{split}
\ee
\end{widetext}
where
\be
\begin{split}
f(p)=&-\frac{2}{E\gamma^2}\bigg\{\sqrt{1+m_\LCperp^2\gamma^2}\,\text{arctan}\left[\frac{1}{p}\sqrt{1+m_\LCperp^2\gamma^2}\right] \\
&-\text{arctan}\left[\frac{1}{p}\right]-\gamma p\,\text{arctan}(\gamma)\bigg\} \;.
\end{split}
\ee 
Note that the pair-production exponential in~\eqref{expSauter}, where we considered ${\bf p}={\bf p}'$, can be expressed with the same function as $\exp(2f(p))$. From~\eqref{preSauterCompton} it is immediately obvious that the exponential vanishes in the limit of low photon energy, $\Omega\to0$, as it must since the probability to emit a soft photon is not exponentially suppressed (the saddle-point approximation breaks down in this limit). 

This rather complicated expression simplifies in the high-energy limit, 
\be\label{ComptonPWlimit}
\begin{split}
\lim_{p,\Omega\gg1}&{\bf M}_{\rm Sauter}({\sf k})=\frac{\alpha a_0\{\kappa-1,0,0,1\}}{4r\sqrt{(1+a_0^2)\text{arccot}(a_0)}} \\
&\times\frac{\exp\left\{-\frac{r a_0}{\chi}\left[(1+a_0^2)\text{arccot}(a_0)-a_0\right]\right\}}{\sqrt{(1+a_0^2)\text{arccot}(a_0)-a_0}} \;,
\end{split}
\ee
where $r=(1/s)-1$, $\kappa=(1/s)+s$ and (in the high-energy limit) $\chi:=\sqrt{(Fp)^2}\to E|{\bf p}|$,  and $s\to(|{\bf p}|-\Omega)/|{\bf p}|$. The first component of~\eqref{ComptonPWlimit} (which gives the probability summed over the photon polarization, $\sum_{\rm pol.} P=2\{1,0,0,0\}\cdot{\bf M}$) agrees with Eq.~(93) in~\cite{Dinu:2018efz} for nonlinear Compton scattering in a plane-wave field, where $s:=np'/np$ with $n_\mu$ being proportional to the wave vector of the plane wave\footnote{Eq.~(93) in~\cite{Dinu:2018efz} actually gives the probability for a large class of symmetric fields, of which the Sauter pulse is one example. In comparing with the lightfront longitudinal spectrum in~\cite{Dinu:2018efz} one should also note that there the longitudinal momenta have been normalized to the initial momentum, so in this case we should write $\ud{\sf k}=p\ud\hat{\sf k}$ and include this extra factor of $p$ into the spectrum.}. To check the other components in~\eqref{ComptonPWlimit} we perform the integrals in Eq.~(24), (25) and (28) in~\cite{Dinu:2019pau} with the saddle-point method as described after~\eqref{MBWplaneWave}. The result agrees exactly with~\eqref{ComptonPWlimit}.

\subsection{Instanton}

\begin{figure}
\includegraphics[width=\linewidth]{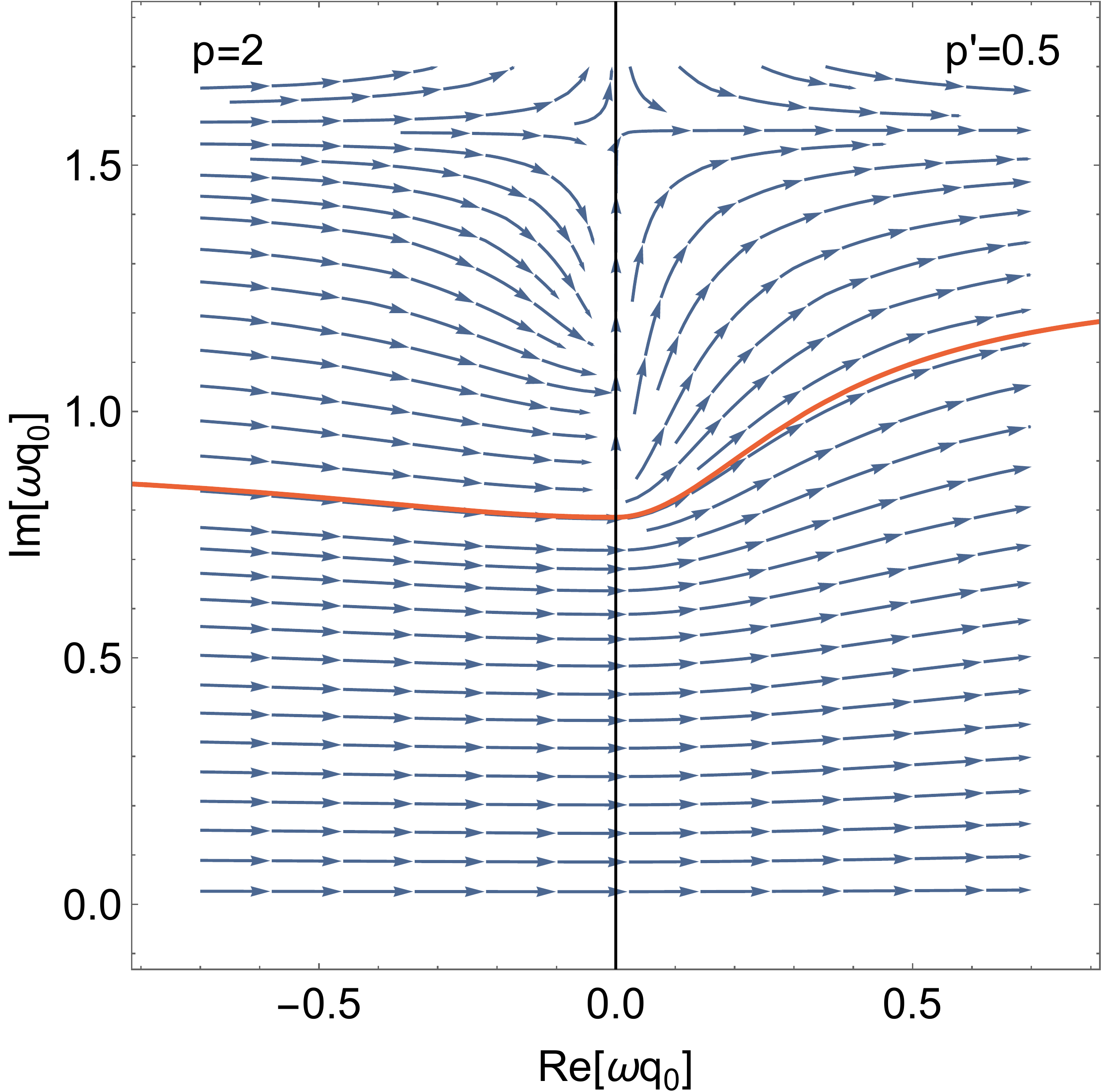}
\caption{Instanton for nonlinear Compton scattering. The momenta are perpendicular to the electric field, $p_1=p$, $p_2=p_3=0$ and $p'_1=p'$, $p'_2=p'_3=0$, and $\gamma=1$. The stream lines show $\sqrt{m_\LCperp^2+A^2(\omega t)}$ for $\text{Re }t<0$ and $\sqrt{{m'_\LCperp}^2+A^2(\omega t)}$ for $\text{Re }t>0$, corresponding to before and after the photon emission.}
\label{streamPlotSauterComptonFig}
\end{figure}

As already mentioned for the pair-production case, while we have for these fields been able to obtain the final results for the probability without actually having to find the instantons explicitly, it is nevertheless useful to consider these instantons since they serve as a starting point for more complicated fields, for which one has to find the instantons explicitly. We consider a Sauter pulse for simplicity. Instead of $\tau$, we parameterize the first part of the instanton (before photon emission) with
\be
U=\sqrt{1+(\gamma m_\LCperp)^2}ET(\tau-\sigma) \;,
\ee
and the second part (after emission) with
\be
U'=\sqrt{1+(\gamma m'_\LCperp)^2}ET(\tau-\sigma) \;.
\ee
For $U<0$ we find
\be\label{q0Compton}
\begin{split}
\omega q^0(U)=&\text{arcsinh}\bigg[\frac{\gamma m_\LCperp}{\sqrt{1+(\gamma m_\LCperp)^2}}\sinh\bigg(U\\
&+i\text{arcsin}\left[\frac{\sqrt{1+(\gamma m_\LCperp)^2}}{m_\LCperp\sqrt{1+\gamma^2}}\right]\bigg)\bigg]
\end{split}
\ee
and
\be\label{q3Compton}
\begin{split}
\omega q^3(U)=&\frac{1}{\sqrt{1+(\gamma m_\LCperp)^2}}\bigg(\text{arcsinh}\bigg[\gamma m_\LCperp\cosh\bigg(U\\
&+i\text{arcsin}\left[\frac{\sqrt{1+(\gamma m_\LCperp)^2}}{m_\LCperp\sqrt{1+\gamma^2}}\right]\bigg)\bigg]\\
&-\text{arcsinh}\left[\frac{\gamma p}{\sqrt{1+\gamma^2}}\right]\bigg) \;.
\end{split}
\ee
The second part ($U'>0$) is obtained from~\eqref{q0Compton} and~\eqref{q3Compton} by simply replacing $U\to U'$ and $p\to p'$. By inserting this solution into~\eqref{finalExpGeneralField} we find agreement with the exponent in~\eqref{preSauterCompton}, where the integral from $U=-\infty$ to $U=0$ gives $\exp[-f(p)]$ and the integral from $U'=0$ to $U'=\infty$ gives $\exp[f(p')]$. The instanton is illustrated in Fig.~\ref{streamPlotSauterComptonFig}.

\section{Conclusions}

We have shown how to use worldline instantons on the amplitude level rather than the probability level, which has been the focus of previous studies\footnote{Recall that the effective action or the photon polarization tensor are effectively probability level when their imaginary parts are used to consider pair production.}. The worldline instantons are then open lines rather than closed loops. We have shown how to amputate the amplitude with respect to the asymptotic fermion states, and how to calculate both the exponential part of the probability as well as how to use the Gelfand-Yaglom method to obtain the full pre-exponential part. 

Working with instantons on the amplitude level is a new approach even if one only considers spontaneous pair production. We expect that it can be a useful alternative for obtaining the momentum spectrum. But we have also shown how to use the instanton formalism for nonlinear Breit-Wheeler pair production and nonlinear Compton scattering. For such processes the instanton has a kink, i.e. a discontinuous velocity, at the point where the (incoherent, high-energy) photon is absorbed or emitted. We have found that the Gelfand-Yaglom method still works. While the probability of nonlinear Breit-Wheeler can also be obtained from probability-level instantons, i.e. by working with the imaginary part of the photon polarization tensor, the probability of nonlinear Compton scattering is not the imaginary part of some closed fermion loop. Thus, our amplitude-level approach allows us to consider more general processes. 

While we have focused on time-dependent electric fields, we have shown that our results for nonlinear Breit-Wheeler and Compton reduce to the corresponding results in a plane-wave background field in the high-energy limit, which were obtained using the standard approach, i.e. with Volkov's wave functions (solution to the Dirac equation) in the Furry picture. The Volkov solutions are very simple, but as soon as one considers other backgrounds than plane waves the solutions to the Dirac equation become much more complicated. But for a time-dependent electric field, the WKB approximations of the wave functions are still simple. We have used these WKB approximations to check our results obtained with the instanton formalism. It is fair to say that for such simple background fields the WKB approach often involves shorter calculations. One reason for this is simply that in the worldline approach we start at a higher level, i.e. the starting point comes before the LSZ reduction, while in the WKB approach this has essentially already been done. One aspect of the worldline instanton approach is that it gives nice semiclassical illustrations of particle trajectories. On the probability level these are complex loops, but on the amplitude level the complex, tunneling segments of the instanton at finite times are connected with asymptotic ends that can be easier to interpret as actual particle trajectories of fermions before and after going through the background field.

However, the real advantage of the worldline instanton formalism comes when one goes beyond simple fields to more realistic space-time dependent fields. This has been demonstrated in~\cite{Schneider:2018huk}, where a numerical instanton code was developed which allows one to consider general fields for spontaneous pair production. For example, the code was applied to an e-dipole pulse~\cite{Gonoskov:2013ada}, which is an exact solution to the Maxwell's equation with finite length in all four space-time coordinates. After having laid the groundwork with the present paper, in the future we plan to develop a code similar to that in~\cite{Schneider:2018huk} but for processes such as nonlinear Breit-Wheeler and Compton.    

Another future application of this instanton approach would be to apply it to obtain the probability of an electron tunneling through a classically forbidden field region, with a space-time dependent electromagnetic field (see~\cite{Kohlfurst:2021dfk} for a recent study of dynamically assisted tunneling) including relativistic effects, or to tunneling in a space-time dependent field assisted by high-energy incoherent photons.

\acknowledgements

We thank Naser Ahmadiniaz for discussions.
G.~T. is supported by the Swedish Research Council, contract 2020-04327.

\appendix

\section{WKB approach}

In this appendix we will briefly explain how to calculate e.g. nonlinear Breit-Wheeler or Compton scattering using the WKB approach.

We treat the photon field as in~\cite{ItzyksonZuber}
\be
A_\mu(x)=\int\frac{\ud^3l}{(2\pi)^32l_0}a_\mu(l) e^{-ilx}+a_\mu^\dagger(l)e^{ilx} \;,
\ee
with
\be
[a_\mu(l),a_\nu^\dagger(l')]=-2l_0(2\pi)^3\delta^3(l'-l)g_{\mu\nu} \;.
\ee
The initial state $|\text{in}\rangle$ contains a photon described by a wave packet $f(k)$ and polarization vector $\epsilon_\mu$,
\be
|\text{in}\rangle=\int\frac{\ud^3k}{(2\pi)^32k_0}f(k)\epsilon(k)a(k)|0\rangle  \;,
\ee
where the normalization $\langle\text{in}|\text{in}\rangle=1$ implies
\be
 \int\frac{\ud^3k}{(2\pi)^32k_0}|f(k)|^2=1 \;.
\ee
The wave packet is sharply peaked. Note that we are using a different normalization for the mode operators of the Dirac and the photon field. For the Dirac field we follow~\cite{Torgrimsson:2017pzs,Hebenstreit:2011pm,Hebenstreit:2010vz,Kluger:1992gb}, which in particular means no factors of $2p_0$ in the integration meassure and the commutation relations. The pair production probability is thus given by
\be
\begin{split}
P&=\int\frac{\ud^3p}{(2\pi)^3}\frac{\ud^3p'}{(2\pi)^3}|\langle0|b_{s',{\bf p}'}^{\rm out}a_{s,{\bf p}}^{\rm out}\mathcal{U}|\text{in}\rangle|^2 \\
&=\frac{e^2}{2\Omega}\int\frac{\ud^3p}{(2\pi)^3}\left|M\right|^2 \;,
\end{split}
\ee
where $\Omega=k_0$, $\mathcal{U}$ is the time evolution operator and
\be\label{PsiGammaPsi}
\begin{split}
&(2\pi)^3\delta^3({\bf p}+{\bf p'}-{\bf k})M:=\langle0|b_{s',{\bf p}'}^{\rm out}a_{s,{\bf p}}^{\rm out}U\epsilon(k)a^\dagger(k)|0\rangle \\
&=\int\ud^4x\bar{\psi}^{(+)}\slashed{\epsilon}e^{-ikx}\psi^{(-)} \;.
\end{split}
\ee
We are only interested here in the saddle-point regime. In this regime we could simply replace the exact wave functions $\psi^{(+/-)}$ with the WKB approximations ${\sf U},{\sf V}$ in~\eqref{UandV}. The spatial integrals are trivial, $\int\ud^3x e^{i(p+p'-k)_jx^j}=(2\pi)^3\delta^3({\bf p}+{\bf p'}-{\bf k})$. The time integral can be performed with the saddle-point method. The saddle point $t_s$ is determined by
\be\label{WKBsaddleBW}
\sqrt{1+({\bf p}'{\bf A}(t_s))^2}+\sqrt{1+({\bf p}-{\bf A}(t_s))^2}\overset{!}{=}\Omega \;,
\ee
Comparing~\eqref{WKBsaddleBW} with~\eqref{eqfortildet} we see that the saddle point for the time integral in the WKB approach coincides with the point $\tilde{t}$ on the instanton where the photon is absorbed.

\section{Prefactor for nonlinear Breit-Wheeler pair production in a constant electric field}\label{prefactor constant field}

In this section we will calculate the prefactor of the probability of nonlinear Breit-Wheeler in a constant field using the worldline formalism. This is an example where the asymptotic fermion states are nontrivial, in contrast to the time-dependent fields we considered in Sec.~\ref{preBWsection} for which the asymptotic states are just plane waves. 
For a general field one can use the instanton method to perform the path integral, but for a constant field $A_3(t)=Et$ this is the same as performing the path integral exactly. So, we start by making the shift and redefinition in~\eqref{shiftRedifinition},
where the Lorentz-force equation simplifies to
\be\label{instantonEqConstant}
\begin{split}
\ddot{q}^0&=T(E\dot{q}^3+J^0) \\
\ddot{q}^3&=T(E\dot{q}^0+J^3) \\
\ddot{q}^\LCperp&=TJ^\LCperp
\end{split}
\ee
with $q^\mu(0)=x'^\mu$ and $q^\mu(1)=x^\mu$. Since we already know that all the nontrivial functional behaviour can be obtained with a perpendicular photon, and since this also gives the maximum probability, we will for simplicity set $k^3=0$. After this shift the path integral becomes (up to an irrelevant phase)
\be\label{pathQuad}
\begin{split}
&\int_{\delta q(0)=0}^{\delta q(1)=0}\mathcal{D}\delta q\exp\left\{-i\int_0^1\frac{\delta\dot{q}^2}{2T}+E\delta q^0\delta\dot{q}^3\right\} \\
&=\frac{1}{(2\pi T)^2}\frac{ET/2}{\sinh(ET/2)} \;.
\end{split}
\ee
This is the same as without the photon.

It is straightforward to solve~\eqref{instantonEqConstant} by first solving in the regions $0<\tau<\sigma$ and $\sigma<\tau<1$ separately and then matching the two parts at $\tau=\sigma$ with $q^\mu(\sigma+\delta)=q^\mu(\sigma-\delta)$ and $\dot{q}^\mu(\sigma+\delta)-\dot{q}^\mu(\sigma-\delta)=Tk^\mu$, where $\delta>0$ and $\delta\to0$. We then plug this instanton solution into the exponent and perform the remaining, ordinary integrals. We change variables to $\varphi^j=(x+x')^j/2$ and $\theta^j=(x-x')^j$. The $\varphi^j$ integral gives $(2\pi)^3\delta^3({\bf p}+{\bf p}'-{\bf k})$. At this point we can take the $t$ derivative, coming from $(i\slashed{\mathcal{D}}_x+m)$, but in the end it turns out that in the asymptotic limit $t\to\infty$ this is the same as the asymptotic limit with $(\slashed{\pi}(t)+m)$. We can now put $t'=t$, and then we shift $Et\to Et+p_3$ which removes $p_3$ from the entire expression. We perform the $\theta^j$ integrals with the saddle-point method. All this is done without taking $t\to\infty$. For the remaining $\sigma$ and $T$ integrals it helps to anticipate their scaling with respect to $t$ before finding the saddle points. We anticipate that $T\to\infty$ as $t\to\infty$, because it takes an infinite proper time to travel from two infinitely separated space-time points. Contrast this with the instanton calculation for the imaginary part of the effective action (i.e. for a closed fermion loop), where $T$ is finite. At this point the exponential part of the integrand for these variables is given by
\be\label{intermediateExpConstant}
\begin{split}
\exp&\bigg\{-\frac{i}{4E}\bigg[2ET(m_\LCperp^2+\sigma k_\LCperp(k-2p)_\LCperp)-\Omega^2\coth\left[\frac{ET}{2}\right]\\
&+\left(2Et_0+\Omega\frac{\cosh\left[ET\left(\sigma-\frac{1}{2}\right)\right]}{\sinh[ET/2]}\right)^2\tanh\left[\frac{ET}{2}\right]\bigg]\bigg\} \;.
\end{split}
\ee  
When performing integrals with the saddle-point method one has to note the scaling of the integration variables with respect to $E$. Here $ET\sim E^0$ and $Eq\sim E^0$. From the above exponential we see that in order to take the asymptotic limit we can change variables from $\sigma\to\frac{1}{2}+\frac{V}{ET}$ to $V$ and from $ET=2\ln\left[\frac{Et}{\Omega X}\right]$ to $X$. After this change of variables we can take the limit $t\to\infty$ of the integrand. The divergent terms from~\eqref{intermediateExpConstant} cancel against the terms in the exponential part of the asymptotic states~\eqref{UandV},
\be
\begin{split}
&i\int^t\pi_p\to\frac{i}{2E}[(Et_0)^2+m_\LCperp^2\ln(Et_0)] \\
&+\text{constant imaginary terms} \;.
\end{split}
\ee
The total exponential becomes
\be
\begin{split}
&\exp\bigg\{-\frac{i}{2E}\Big[k_\LCperp(k-2p)_\LCperp V+4\Omega^2(X\cosh(V)-X^2) \\
&\hspace{2cm}-(m_\LCperp^2+{m'}_\LCperp^2)\ln(X)\Big]\bigg\} \;.
\end{split}
\ee
It is now much easier to perform these integrals with the saddle-point method than if we had done so before changing variables and taking the asymptotic limit. For $V$ we have a saddle point at
\be
V=\text{arcsinh}\left[-\frac{1}{4X}\frac{k_\LCperp(k-2p)_\LCperp}{\Omega^2}\right] \;.
\ee

For simplicity, and in order to compare with~\cite{Dunne:2009gi}, we consider the total probability, i.e. we perform the momentum integrals. For a constant field the longitudinal momentum integral has a constant integrand and gives a volume factor $\int\ud p_3=EV_0$, which is a standard relation. For the perpendicular integrals we have a saddle point at $p_\LCperp=k_\LCperp/2$. It is convenient to expand the exponential to quadratic order around this saddle point already on the amplitude level and before performing the $X$ integral, and afterwards we use $p_\LCperp$ for the saddle point value $p_\LCperp=k_\LCperp/2$ rather than the integration variables. After this we find a simpler exponential and a saddle point at $X=(i+p)/(4p)$, where $p=|{\bf p}|$. 
At the end we find
\be
P\sim\dots\exp\left\{-\frac{2}{E}\left([1+p^2]\text{arctan}\left[\frac{1}{p}\right]-p\right)\right\} \;,
\ee
which agrees with~\cite{Dunne:2009gi}. To obtain the rest of the prefactor we just have to calculate the quadratic variation around each of the above saddle points and collect the contributions from the corresponding Gaussian integrals. 

For the spin and polarization dependent parts of the prefactor we just have to replace the integration variables with their saddle points. For the contribution from the asymptotic electron state we have
\be
\lim_{t\to\infty}\frac{(\slashed{\pi}+1)R}{\sqrt{2\pi_0(\pi_0+\pi_3)}}=\frac{1}{m_\LCperp}(p_\LCperp\gamma^\LCperp+1)R
\ee  
and for the positron (note different momentum)
\be
\lim_{t\to\infty}\frac{(-\pi_0\gamma^0+\pi_j\gamma^j+1+1)R}{\sqrt{2\pi_0(\pi_0-\pi_3)}}\to-\gamma^0R \;.
\ee
We should take the linear part in $\epsilon$. This can come from either $\epsilon\dot{q}$ or from $\slashed{k}\slashed{\epsilon}$; we call these the scalar and spinor parts, respectively. A potential term coming from making the shift $A_\mu\to A_\mu+\epsilon_\mu e^{-ikx}$ in $(i\slashed{\mathcal{D}}_x+m)$ does not contribute. For the scalar part we have
\be\label{expgammagamma}
\begin{split}
&\mathcal{P}\exp\left\{-i\frac{T}{4}\int_0^1\sigma^{\mu\nu}F_{\mu\nu}\right\}R \\
&=\exp\left[\frac{ET}{2}\gamma^0\gamma^3\right]R=\exp\left[\frac{ET}{2}\right]R
\end{split}
\ee
and then
\be
\begin{split}
&\frac{1}{m_\LCperp}\bar{R}_r(p_\LCperp\gamma^\LCperp+1)\gamma^0(i\slashed{\mathcal{D}}_x+m)R_{r'} \\
&\to\frac{1}{m_\LCperp}\bar{R}_r(p_\LCperp\gamma^\LCperp+1)\gamma^0(p_\LCperp\gamma^\LCperp+1)R_{r'}=m_\LCperp R_r^\dagger R_{r'} \;.
\end{split}
\ee
Inserting the saddle points gives $-i\epsilon\dot{q}(\sigma)\to\epsilon_3 T$. Hence, the scalar part is only nonzero (at this leading order in $E$) if the photon polarization has a nonzero longitudinal component, but vanishes for perpendicular polarization. Since we should take the limit $t\to\infty$, it is important to keep track of how different terms scale. For this scalar part we have, with the saddle point for $T$, 
\be\label{scalarScalingt}
T\exp\left[\frac{ET}{2}\right]\sim\mathcal{O}(t\ln t) \;.
\ee

For the spinor part we have
\be
\begin{split}
&\mathcal{P}\exp\left\{\int_0^1\frac{ET}{2}\gamma^0\gamma^3-\frac{iT}{2}\slashed{k}\slashed{\epsilon}e^{-ikq}\right\} \\
&\to\int_0^1\ud\sigma\exp\left\{\frac{ET}{2}(1-\sigma)\gamma^0\gamma^3\right\} \\
&\times\left(-\frac{iT}{2}\slashed{k}\slashed{\epsilon}e^{-ikq}\right)\exp\left\{\frac{ET}{2}\sigma\gamma^0\gamma^3\right\} \;.
\end{split}
\ee
The part to the right of $\slashed{k}\slashed{\epsilon}$ again simplifies as in~\eqref{expgammagamma}. The part to the left of $\slashed{k}\slashed{\epsilon}$ simplifies to
\be\label{leftPart}
\begin{split}
&\frac{1}{m_\LCperp}R_r^\dagger\bigg[(\partial_t-\pi_3)\gamma^0(p_\LCperp\gamma^\LCperp+1)\exp\left\{-\frac{ET}{2}(1-\sigma)\right\} \\
&+m_\LCperp^2\exp\left\{\frac{ET}{2}(1-\sigma)\right\}\bigg] \;.
\end{split}
\ee
With $\partial_t-\pi_3\to2Et$ and $\sigma\to1/2$ we have 
\be\label{leftPart2}
\begin{split}
&\eqref{leftPart}\exp\left\{\frac{ET}{2}\sigma\right\}T \\
&\to\frac{1}{m_\LCperp}R^\dagger\left[2Et\gamma^0(p_\LCperp\gamma^\LCperp+1)+m_\LCperp^2\exp\left[\frac{ET}{2}\right]\right]T \;.
\end{split}
\ee
Since $ET=2\ln\left[\frac{Et}{\Omega X}\right]$, all the terms in~\eqref{leftPart2} are $\mathcal{O}(t\ln t)$, same scaling as the scalar contribution~\eqref{scalarScalingt}. As $t\to\infty$ the only other $t$ dependent contribution comes from $\eqref{pathQuad}\sim\mathcal{O}(1/[t\ln t])$, which hence gives a finite limit as $t\to\infty$. 

We have used $\gamma^0\gamma^3R=R$. We can choose a basis with
\be
i\gamma^1\gamma^2R_n=(-1)^nR_n \qquad n=1,2 
\ee
to calculate the rest of the spinor part. We sum for simplicity over the spins. 

We finally obtain $P={\bf N}\cdot{\bf M}$ with
\be\label{PparaperpConstant}
\begin{split}
{\bf M}&=\frac{\alpha E V_0}{4\Omega p m_\LCperp}\frac{\exp\left\{-\frac{2}{E}\left(m_\LCperp^2\text{arccot}(p)-p\right)\right\}}{\sqrt{\text{arccot}(p)\left(m_\LCperp^2\text{arccot}(p)-p\right)}} \\
&\times\{1+3p^2,0,0,1-p^2\}  \;.
\end{split}
\ee
For parallel and perpendicular polarization, ${\bf N}=\{1,0,0,1\}$ and ${\bf N}=\{1,0,0,-1\}$, ${\bf N}\cdot{\bf M}$ agrees with Eq.~(5)-(8) in~\cite{Dunne:2009gi}.

\end{document}